\documentclass[twocolumn]{aastex631}

\usepackage{xspace}
\usepackage{multirow}

\newcommand{\kepler}{\textit{Kepler}\xspace}
\newcommand{\gaia}{\textit{Gaia}\xspace}
\newcommand{\tess}{\textit{TESS}\xspace}
\newcommand{\vsini}{$v\sin{i}$\xspace}
\newcommand{\Prot}{$P_\mathrm{rot}$\xspace}
\newcommand{\Pcyc}{$P_\mathrm{cyc}$\xspace}
\newcommand{\RHK}{$R'_\mathrm{HK}$\xspace}
\newcommand{\logRHK}{$\log{R'_\mathrm{HK}}$\xspace}
\newcommand{\Teff}{$T_\mathrm{eff}$\xspace}
\newcommand{\CaII}{\ion{Ca}{2}\xspace}
\newcommand{\Sindex}{$S$-index\xspace}

\begin{document}

\title{HWO Target Stars and Systems: Activity and Rotation Catalog (ARC) of Potential Target Stars for the Habitable Worlds Observatory}

\author[0000-0002-3551-279X]{Tara Fetherolf}
\altaffiliation{NASA Postdoctoral Program Fellow}
\affiliation{Department of Earth and Planetary Sciences, University of California, Riverside, CA 92521, USA}
\email{Tara.Fetherolf@gmail.com}

\author[0000-0002-5463-9980]{Arvind F. Gupta}
\affiliation{U.S. National Science Foundation National Optical-Infrared Astronomy Research Laboratory, Tucson, AZ 85719, USA}

\author[0000-0003-4150-841X]{Elisabeth R. Newton}
\affiliation{Department of Physics and Astronomy, Dartmouth College, Hanover, NH 03755, USA}

\author[0000-0001-9546-9044]{Andrea P. Buccino}
\affiliation{Instituto de Astronom\'{i}a y F\'{i}sica del Espacio (IAFE, CONICET-UBA), Buenos Aires, Argentina}
\affiliation{Departamento de F\'{i}sica, Facultad de Ciencias Exactas y Naturales, Universidad de Buenos Aires, Buenos Aires, Argentina}

\author[0000-0002-0040-6815]{Jennifer A. Burt}
\affiliation{Jet Propulsion Laboratory, California Institute of Technology, Pasadena, CA 91109, USA}

\author[0000-0002-7349-1387]{Jos\'{e} A. Caballero}
\affiliation{Centro de Astrobiolog\'{i}a, CSIC-INTA, ESAC Campus, Camino Bajo del Castillo s/n, 28692 Villanueva de la Ca\~{n}ada, Madrid, Spain}

\author[0009-0006-9244-3707]{Sebasti\'{a}n Carrazco-Gaxiola}
\affiliation{Department of Physics and Astronomy, Georgia State University, Atlanta, GA 30302-4106, USA}
\affiliation{RECONS Institute, Chambersburg, PA 17201, USA}

\author[0000-0003-4615-8746]{Mariela C. Vieytes}
\affiliation{Instituto de Astronom\'{i}a y F\'{i}sica del Espacio (IAFE, CONICET-UBA), Buenos Aires, Argentina}
\affiliation{Departamento de Ciencia y Tecnolog\'{i}a, UNTREF, Buenos Aires, Argentina}

\author[0000-0003-0595-5132]{Natalie R. Hinkel}
\affiliation{Department of Physics \& Astronomy, Louisiana State University, 202 Nicholson Hall Baton Rouge, LA 70803, USA}

\author[0000-0003-2008-1488]{Eric E. Mamajek}
\affiliation{Jet Propulsion Laboratory, California Institute of Technology, Pasadena, CA 91109, USA}

\begin{abstract}
A major goal of the Habitable Worlds Observatory (HWO) is to precisely characterize exoplanets and their atmospheres. However, magnetic activity from an exoplanet's host star can complicate measurements of both the stellar and planetary properties, and stellar activity can be an important factor in our interpretation of the evolutionary history of an exoplanet. In this work, we assess the extent to which magnetic activity has been characterized for potential HWO target stars by collating archival measurements of relevant observables as published in a broad range of photometric and spectroscopic datasets. We describe our data collection strategy, provide an overview of currently known activity and rotation properties in the Activity and Rotation Catalog (ARC) for potential HWO target stars, and briefly review known relationships between stellar inclination, rotation, activity, and age. Overall, we find that stellar activity (\Sindex and \RHK) and rotation (\vsini and \Prot) properties have been measured for at least 70\% systems that are currently of high interest as potential HWO atmospheric characterization targets. However, stellar activity is temporal in nature, such that activity properties should be regularly monitored in order to remain up-to-date for informing future observations. In particular, we find that stellar activity cycles are measured for fewer than 20\% of high interest potential HWO target stars. Measuring a star's activity cycle is critical for anticipating times when higher levels of magnetic activity may occur during planned HWO observations, which may interfere with measuring precise exoplanet atmospheric characteristics. 

\end{abstract}



\section{Introduction} 
\label{sec:intro}
The Habitable Worlds Observatory (HWO) is a NASA flagship space mission concept currently in early development \citep{Feinberg2024,Feinberg2026}. One of the primary goals of the HWO mission is to directly image Earth-sized exoplanets in the habitable zones of Sun-like stars and characterize their atmospheres \citep{Arney2026}. Only a limited selection of stars will be able to be observed due to the difficulty in obtaining sufficient planet-star contrast at close angular separations. Therefore, potential targets need to be carefully selected in order to maximize the scientific outcomes of the mission \citep{Tuchow2025}. It is well known that measurements of most exoplanet properties of interest are only as precise as the measurements of their host star properties \citep[e.g.,][]{Kane2014, Ciardi2015}. Therefore, it is vital to carefully and thoroughly characterize stars in order to assess whether they are optimal targets for HWO and similar exoplanet missions. In particular, stellar magnetic activity is a major barrier towards measuring precise stellar and exoplanet properties. Many previous studies have prioritized searching for and characterizing exoplanets around quiet stars \citep[e.g.,][]{Gupta2021, Brewer2020, Simpson2023, Fetherolf2026}, but stars shift from being quiescent to active over the course of decades such that a quiet star may become active by the time of HWO observations. However, accurate exoplanet properties can still be obtained if stellar activity and variability is well-characterized or mitigated during observations \citep[e.g.,][]{Dumusque2014, Vanderburg2016, Oshagh2017, Kaplan-Lipkin2022, Hara2023, Dorval2024, Rackham2024, Klein2025}. The Activity and Rotation Task Group was formed within the Target Stars and Systems (TSS) sub-working group\footnote{The TSS sub-working group was established as part of the Living Worlds Working Group designated by the HWO Science, Technology, Architecture Review Team (START).} in order to collect existing measurements of stellar activity and rotation rates, and identify gaps in knowledge for stars that may potentially be selected for observations with the future HWO mission. This work is representative of the efforts of the TSS Activity and Rotation Task Group.

Observed signals from stellar activity can hide planetary signals or be mistaken for being planetary in nature \citep[e.g.,][]{Henry2002, Robertson2014, Robertson2015, Lubin2021, Simpson2022, Niraula2022}, which can lead to inaccurate measurements of planetary properties, a misunderstanding of system architectures, and a misrepresentation of known exoplanet demographics \citep[e.g.,][]{Kane2014, Ciardi2015, Osborne2025}. Stellar activity will inevitably affect our ability to accurately measure and interpret exoplanet atmospheric spectra. 
Observational evidence has shown that the equivalent widths of numerous metallic lines increase systematically with activity \citep{Wise2018, Spina2020}, while HARPS-N observations of the Sun-as-a-star reveal that this variability is highly species-dependent, with some lines strengthening while others weaken in response to varying activity levels \citep{Dravins2024}. Recent non-LTE modeling of G2 dwarf stars across varying levels of chromospheric activity has identified that these variations are primarily driven by changes in line contribution functions \citep{Vieytes2025}. Building upon these models, unsupervised machine learning classifications have further revealed that line sensitivity is strongly correlated with atomic parameters; specifically, approximately 11\% of transitions—predominantly violet \ion{Fe}{1} lines with low excitation energies—are highly susceptible to chromospheric heating due to Saha ionization effects (Peralta \& Vieytes, in prep.).
Just as variations in host star flux due to magnetic activity are known to complicate transmission spectroscopy measurements \citep{Zellem2017, Rackham2018}, these variations will similarly affect reflected light spectroscopy for directly imaged planets as they will alter both the incident flux spectrum and the background stellar spectrum. Stellar activity and variability could also impact the success of starlight suppression techniques when using a coronograph, given that coronagraphic imaging is often used to observe coronal mass ejection events from the Sun \citep[e.g.,][]{Yashiro2004}. For HWO, we will therefore need strong constraints on the baseline stellar activity level at the time of each observation and on how the activity varies over the course of the observation(s). Understanding how activity levels vary on both short and long timescales will be integral to designing a successful survey strategy. 

All stars go through a phase of increased magnetic field activity at some point in their evolution \citep[e.g.,][]{Vidotto2014, Getman2023, Isik2023}, and this activity is likely to have an influence on the size and chemical makeup of exoplanet atmospheres. Close-orbiting planets, for example, can experience significant atmospheric escape driven by high-energy radiation \citep[e.g.,][]{Sanz-Forcada2011, Owen2019, Bourrier2020, Saidel2026}, which can result in a planet having little to no atmosphere. Activity levels and stellar rotation periods are also commonly used as proxies to estimate stellar ages \citep[e.g.,][]{Irwin2009, Meibom2009, vanSaders2013, Pinsonneault2012, Angus2019, Douglas2024}. The age of a star in turn informs constraints on the evolutionary state of planetary atmospheres and surface chemistry. Recently, there have been a handful of planets that have been observed to have either insignificant or opaque atmospheres \citep{Kreidberg2019, Crossfield2022, Greene2023, Zieba2023, MeierValdes2025, Piaulet-Ghorayeb2025, Fortune2025, Xue2025, Allen2025, Wachiraphan2025, Bennett2025, Fisher2026}. However, due to the uncertainty in the ages of the host stars, it is not clear if the planetary atmospheres were potentially lost due to the geological age of the planet (e.g., loss the primary atmosphere before the formation of the secondary atmosphere) or past stellar activity. Reliable measurements of stellar rotation periods and projected rotational velocities can be used to calculate the stellar inclination and predict planetary system orientations. Overall, characterizing stellar activity levels for HWO target stars is important for understanding the evolutionary history of planetary systems, providing context for measured exoplanet atmospheres, and gaining a complete picture of planetary system architectures.

Measuring the activity of potential HWO target stars now is important for understanding the typical activity variations that may take place over the timescale of the observations (several hours in a single visit), but also for understanding how the star varies over the course of its magnetic activity cycle. Precisely measuring exoplanet properties is easier around quiet stars, but stars experience different levels of activity over many-year timescales \citep[e.g.,][]{Baliunas1995, Baum2022, Isaacson2024, Ramsay2024}. A star that is considered inactive based on current observations may enter a period of maximum activity during planned HWO observations, or vice versa. Being informed about stellar activity cycles will allow for prioritization of stars that exhibit low activity levels during the time of planned observations. Furthermore, gaining an overall better understanding of stellar activity will enhance the development of stellar noise mitigation techniques that can be applied to observational strategies or data analyses. Measuring stellar activity cycles requires monitoring magnetic activity levels (i.e., flux amplitude variations, flare rates, and average activity indicators) over long time baselines, which means potential HWO stars need to start having their magnetic activity levels monitored now---especially if these observations do not already exist. 

In this work we have collected and ranked existing measurements of stellar rotation and activity parameters for potential HWO targets. In addition to collating literature measurements into a single Activity and Rotation Catalog (ARC), we identified knowledge gaps that must be filled in preparation for HWO and other future direct imaging missions. In Section~\ref{sec:current_best}, we briefly review the current state-of-the-art methods for monitoring stellar activity and rotation using spectroscopy and photometry. Development of the ARC, the major product from the HWO TSS Activity and Rotation Task Group presented in this work, is described in Section~\ref{sec:methods}. The contents and completeness of the ARC are provided in Section~\ref{sec:catalog}. We summarize the major implications based on relationships between activity and rotation properties formed from the contents of the ARC in Section~\ref{sec:discussion}. We then discuss known efforts for ongoing and future stellar activity monitoring in Section~\ref{sec:future}. Finally, in Section~\ref{sec:summary}, we summarize our findings and provide additional recommendations for future stellar activity monitoring that will best support HWO and other future exoplanet missions.


\section{Current State-of-the-Art Activity Monitoring}
\label{sec:current_best}
Currently, stellar magnetic activity is monitored both spectroscopically and photometrically. Spectroscopic methods generally measure the average chromospheric magnetic activity levels at the time of the observations, and can also be used to place constraints on stellar rotations based on the measured rotational line broadening. Photometric methods show how the summation of all contributions of the star’s light in a given bandpass changes over time, with significant flux variations typically being caused by starspots on the photosphere or magnetic flare events. Most measurements of stellar activity occur on relatively short timescales (few hours to few weeks), but a small sample of stars ($\lesssim$100) have had their magnetic activity monitored for decades, allowing for measurements of their magnetic activity cycles. The following subsections describe in further detail the current state-of-the-art measurements for monitoring stellar magnetic activity using spectroscopic and photometric methods. 


\subsection{Spectroscopic Activity Monitoring}
Spectroscopic observations typically trace stellar activity levels via measurement of spectral lines that are known to be particularly sensitive to chromospheric activity. Monitoring of these lines on both short and long timescales can provide information about a star’s typical activity level, the stellar rotation period, and the long-term magnetic activity cycle. A long-running survey conducted at Mount Wilson \citep{Wilson1978} established the use of \CaII H \& K emission core measurements to monitor stellar activity variations, and to date this survey remains the standard for the field. Numerous subsequent activity monitoring programs have made use of this same metric, and variations thereof, to measure activity levels and variability for other stars and to extend the Mount Wilson baseline. Other commonly used activity proxies include H$\alpha$, the Na D doublet, and the \CaII infrared triplet \citep[IRT;][]{Cincunegui2007, Busa2007, Schofer2019}.

Aside from the Mount Wilson survey, much of what we have learned about activity levels from spectroscopic data has come as a byproduct of radial velocity (RV) exoplanet search programs. Because activity-induced surface features such as starspots and faculae produce apparent RV variations, activity can often mask or mimic exoplanet-induced signals \citep{Queloz2001, Borgniet2015}. Astronomers have thus taken care to capture activity proxies such as the \Sindex \citep[][see also Section \ref{sec:s_index}]{Duncan1991} and \RHK \citep[][see also Section \ref{sec:rhk}]{Noyes1984}, which quantify emission in the \CaII H \& K line cores, and spectral line shape metrics such as the full-width at half-maximum (FWHM) and bisector inverse slope of the cross-correlation function (CCF) with which most optical spectrographs extract RVs. This information is used to disentangle apparent RV shifts due to line shape deformations from true, center-of-mass motions. As a result, S-values and activity cycles have thus far been best measured for many Northern hemisphere stars using HIRES data from the California Legacy Survey \citep{Rosenthal2021, Isaacson2024}, and for Southern hemisphere stars using HARPS data \citep{GomesdaSilva2021}. 

With a sufficient volume of data over decades-long baselines, stellar activity cycles can be measured. However, rotation periods are somewhat more difficult to measure with spectroscopic observations. Spectroscopic measurements are not always collected at high enough cadence to resolve the quasi-periodic rotational activity modulation, particularly for relatively inactive stars with weaker variations, which are the main focus of many RV exoplanet searches. Some rotation periods have been measured with data from the Mount Wilson survey \citep[e.g.,][]{Olah2016} and from HARPS \citep{SuarezMascareno2015} and TIGRE \citep{Hempelmann2016, Mittag2017}, but most stars lack spectroscopic rotation period measurements. With just a single high resolution, high S/N spectrum, however, one can estimate the projected rotational velocity, \vsini, by measuring the line broadening. This provides an upper limit on rotation period. Projected rotational velocities have been extracted from several large spectroscopic data sets \citep[e.g.,][]{Schroder2009, Brewer2016}. These values are typically very reliable for fast-rotating stars, but the spectral resolution of most of these data sets is insufficient to measure broadening of less than a few km/s.


\subsection{Photometric Activity Monitoring}
Time-series photometry is generally used to monitor changes in starlight on short timescales (few hours to weeks). Photometric variability in stars can be caused by eclipsing binaries, high-frequency stellar oscillations, transient events, and magnetic field activity. Average stellar flux levels, or in rare cases flare events, can be captured in a single night of ground-based observations. Changes in stellar fluxes that are attributed to rotational modulations from starspots can be captured through repeat visits over days and weeks. 

Before the launch of space-based photometric missions, wide-field ground-based photometric surveys, such as the All-Sky Automated Survey \citep[ASAS;][]{Pojmanski2002}, the Kilodegree Extremely Little Telescope \citep[KELT;][]{Oelkers2018}, the Asteroid Terrestrial-impact Last Alert System \citep[ATLAS;][]{Heinze2018}, and the All-Sky Automated Survey for Supernovae \citep[ASAS-SN;][]{Jayasinghe2018}, have largely led the discovery of variable stars across the sky. While wide-field ground-based surveys generally intend to search for rare transient events, such as supernova, microlensing, and asteroid occultations, they inevitably also detect thousands of stars exhibiting rotational modulations and magnetic activity. Ground-based time-series photometry is also incredibly accessible to amateur astronomers, such that there is a vast record of variable stars that have been reported in the the Variable Star Index \citep{Watson2006}. 

Space-based time-series photometry, including CoRoT \citep{Auvergne2009}, \kepler \citep{Borucki2010}, and \tess \citep{Ricker2015}, revolutionized the study of variable stars by being able to simultaneously monitor tens to hundreds of thousands of stars at high-precision and nearly continuously for time periods of weeks, months, and even years. In particular, the increased sensitivity of space-based photometry enabled detections of lower amplitude stellar variations, such that several \kepler studies found that the fraction of variable stars relative to quiet stars increases with photometric precision \citep[e.g.,][]{Ciardi2011, Basri2011, Briegal2022}. These findings suggest that we are currently observationally limited in identifying variable stars and there are, in fact, many more variable stars than are currently known. 
Furthermore, stellar variability has been found to have an impact on \gaia photometry \citep{GaiaCollaboration2016, GaiaCollaboration2019}, which will need to be corrected when searching for exoplanets with astrometry. As \kepler is no longer operating and \gaia photometric monitoring is sparse, \tess has become the leader in space-based monitoring of stellar variability and activity across the sky \citep[e.g.,][]{Fetherolf2023, Colman2024, Claytor2024}. \tess recently completed 7 years of observations, and will likely continue revisiting each patch of the sky for years to come. Overall, the amount of photometric data available for monitoring stellar activity across the sky is vast and rapidly growing, but the analyses and interpretation of the data available is currently limited by computation speed and manpower. 


\section{Literature Compilation} 
\label{sec:methods}
We have compiled stellar magnetic activity and rotation measurements of potential HWO target stars from the literature in order to assess current knowledge of these properties and identify gaps that need to be filled in preparation for future missions. In this section, we describe the creation of the ARC. The stars selected for the ARC are based on those presented in the HWO TSS 2025 \citep[TSS25;][]{Tuchow2025} list, which is a prioritized target star list based on the HWO Preliminary Input Catalog \citep[HPIC;][]{Tuchow2024}. The TSS25 is a list of potential target stars that may be updated in the future. Existing catalogs of activity and rotation measurements from the literature were compiled and assessed based on the number of stars, the quality and spectral resolution of the observational data, and the spectral types of the stars contained within each catalog. Catalogs containing larger numbers of stars and higher quality observational data were generally prioritized, and for this work catalogs that primarily contained M dwarf stars or \kepler stars were deprioritized since these stars are not anticipated to be high priority HWO targets. The activity properties collected include measurements that quantify the amount of stellar magnetic activity (\Sindex, \RHK, jitter), stellar rotation rates (\vsini, \Prot), and periodicity of magnetic activity cycles (\Pcyc). 

In terms of chromospheric diagnostics, we prioritize the \CaII H\& K doublet due to its long-standing calibration and sensitivity in the low-activity regime characteristic of HWO targets. While H$\alpha$ is a valuable indicator of stellar activity, particularly in more active or later-type stars, it was not collected for this catalog.  First, in Solar-type stars, the H$\alpha$ profile is dominated by a deep photospheric absorption component, meaning that chromospheric activity manifests primarily as a subtle ``filling-in'' of the line core rather than a distinct emission feature. This makes the resulting activity index highly sensitive to the specific integration bandwidth used \citep{GomesdaSilva2022}.  Furthermore, there is plenty of evidence that the correlation between H$\alpha$ emission to the \CaII activity indexes is not straightforward. While these diagnostics may scale together in high-activity regimes, the relationship often weakens, disappears, or even exhibits anti-correlations in quiet, Solar-type stars \citep{Cincunegui2007, GomesdaSilva2014, GomesdaSilva2021, Maldonado2022, IbanezBustos2023}. This decoupling is generally attributed to the different formation heights of the lines within the stellar atmosphere and the varying sensitivity of each species to different magnetic structures, such as the contrasting flux contributions from chromospheric plages versus cool filaments.

Although the  activity indicators \Sindex and \RHK are derived from the same \CaII features, they are employed for different analytical purposes. The primary advantage of the \Sindex is that it can be calculated as a flux ratio directly from non-flux-calibrated spectra, though it requires careful cross-calibration when combining data from different instruments.  This makes the \Sindex the standard for long-term variability and stellar cycle analysis \citep{Baliunas1995, Baum2022}.  In contrast, \RHK provides an absolute physical measurement of chromospheric activity by normalizing the emission to the stellar bolometric luminosity and subtracting the photospheric contribution. This normalization allows for the direct comparison of activity levels across stars of different spectral types and facilitates correlations with fundamental parameters such as rotation and age \citep{Noyes1984}. However, its calculation is inherently model-dependent, as it relies on specific photospheric ``basement'' subtractions and continuum models \citep{Boudreaux2022}. By providing both metrics, we ensure the catalog maintains both historical observational continuity and physical interpretability. 

The stars contained within each literature catalog were matched with the HPIC using any available object identifier (e.g., TIC, HD, Gaia, etc.) or, if needed, positional matching. For each activity property, the literature catalogs providing measured values for that specific property were ranked based on the quality of the measurements. Generally, literature catalogs that utilized higher resolution data or contained hand-curated activity measurements received a higher ranking. Each activity property had an independent ranking of literature catalogs, such that a given catalog receiving a high ranking in one activity property did not guarantee that it would receive a high ranking in another activity property. This treatment was followed since some literature catalogs would analytically convert between activity indicators using the same dataset, rather than obtain independent measurements for each activity indicator. 

In the subsections below, the literature catalogs used for each activity property (\vsini, \Sindex, \RHK, \Prot, \Pcyc, and jitter) are listed in their ranked order, with highest ranking listed first. We provide a brief description of the literature catalog alongside any notes for special treatment of a given catalog, if needed. For each star in the ARC, the value provided (if available) is based on the highest ranking literature catalog for each activity property. Catalogs with multiple activity properties may not necessarily be the highest ranked in each of those properties. Therefore, it is important to note that measurements for a given star are not necessarily collected at the same time and may reflect different levels of magnetic activity. In the case where a catalog contained duplicate entries for a given star, we chose to select the first entry listed in the literature catalog. Throughout the remainder of this work, we refer to the selected highest ranking values that presented in the ARC as the ``adopted'' values. 


\subsection{Projected Rotational Velocity}
\label{sec:vsini}
The projected stellar rotational velocity, \vsini, is determined via measurement of spectral line broadening. Results will thus depend on the specific treatment of other broadening factors. For example, different treatments of microturbulence ($v_\mathrm{mic}$ or $\xi_t$) and macroturbulence ($v_\mathrm{mac}$) often yield different \vsini measurements for a given star, even when using the same input spectra. In addition, the minimum rotational velocity that can be measured depends on the resolving power of the spectrograph with which data were collected. One cannot measure broadening that is much smaller than the size of the resolution element. Values reported in the literature also differ in methodology, with some measurements adopting a forward modeling approach and others using empirical relations or data-driven analysis techniques. A detailed discussion and comparison of the different adjustment methods and assumptions is given in \citet{Blanco-Cuaresma2019}.

For the ARC we adopt \vsini measurements from the following enumerated sources, which are listed in decreasing order of priority. The enumerated sources include an appended instrument specification in cases where the datasets received different rankings. Here, we prioritize measurements at higher spectral resolution. While we place no emphasis on any specific analysis technique with respect to microturbulence or macroturbulence, we do assign a higher priority to catalogs with greater overlap with the HPIC, such that our final output catalog is more self-consistent. We omit references that supply \vsini measurements but do not explain how they were computed, and references that report \vsini values measured from spectra with $R<10$,$000$.
We also exclude measurements of stars that share TIC identifiers, i.e., multiple components of a spectroscopic binary system. 

\vsini may be reported as an upper limit when the combination of \vsini value and instrumental resolution prevents detection. Authors assess and report upper limits in different ways. In this work, upper limits are indicated with the value $-99$; readers should refer to the original source for an understanding of what limits this places on the star's actual rotational velocity. These are selected in the same priority scheme as detections: if a higher rank (typically higher resolution) catalog reports a non-detection, it will not be superceded by a detection from a lower rank (typically lower resolution) catalog.

When error bars were not provided, we estimated them by calculating the standard deviation with respect to a standard, accounting for the error in the latter values. We also identified three instances (two sources from \citealt{Luck2017} and \citealt{Zorec2012}) in which there were clear offsets between these values and others in the ARC. In these cases, we applied a correction derived from the median \vsini\ with respect to a standard. The values and standards used for errors and offsets are indicated in each reference.

\begin{enumerate}

\item \citet{Soto2018}: \vsini measurements were derived from HARPS spectra ($R\sim\,$115,000) for 468 HPIC stars using the SPECIES code. SPECIES \citep{Soto2018} calculates \vsini by fitting the observed line profile to synthetic lines convolved with a rotation kernel, where $v_\mathrm{mac}$ is scaled from solar values for individual lines as described in \citet{dosSantos2016} and $v_\mathrm{mic}$ is computed along with other atmospheric parameters via equivalent width measurements. Error bars of $1.7$~km/s were adopted based on comparison to \citet{CostaSilva2020}. Values of $0$~km/s are marked as non-detections.

\item \citet{CostaSilva2020}: \vsini measurements were derived from HARPS spectra ($R\sim\,$115,000) for 387 HPIC stars by fitting the observed spectra to synthetic spectra generated using MOOG \citep{Sneden1973}. $v_\mathrm{mac}$ was determined for each star via an empirical scaling relation given in \citet{Doyle2014}. $v_\mathrm{mic}$ was held fixed, but the authors do not state the specific values or source of values.

\item \citet{Maldonado2022}: \vsini measurements were determined for 65 HPIC stars using spectra from either HARPS or HARPS-N, both of which have spectra resolution $R\sim\,$115,000. \citet{Maldonado2022} computed \vsini using the Fourier transform method described in \citet{Gray2008}. In the case of duplicate measurements for a single star, we select the measurement with the smaller reported uncertainty.

\item \citet{Rainer2023}: \vsini measurements were derived from HARPS-N spectra ($R\sim\,$115,000) for 50 stars on the HPIC list using an empirical relation between the CCF FWHM and \vsini. $v_\mathrm{mic}$ and $v_\mathrm{mac}$ for the calibrator sample were determined using scaling relations taken from \citet{Adibekyan2012} and \citet{Doyle2014}, respectively. Non-detections are included as such.

\item \citet{Borisov2023} \vsini measurements are provided for 81 HPIC stars using $R\sim\,$80,000 UVES spectra and a custom synthetic spectral fitting algorithm. The treatment of $v_\mathrm{mic}$ and $v_\mathrm{mac}$ is not discussed. Values of $0.165$~km/s with no associated error and are marked as non-detections.

\item \citet{Brewer2016}: \vsini measurements were calculated for 538 stars on the HPIC list using high-S/N template spectra from HIRES ($R\sim\,$70,000). The authors forward model a uniform set of spectral lines in the wavelength range 5164~A--7800~A, assuming $v_\mathrm{mic} = 0.85$~km/s for all stars, and using a $v_\mathrm{mac}$ scaling relation derived from a subset of their target sample. The uncertainty on \vsini is assumed to be $0.5$ km~s$^{-1}$ for all stars. Values equal to $0.1$~km/s are marked as non-detections.

\item \citet{Rice2020}: \vsini measurements were calculated for 362 stars on the HPIC list using archival HIRES spectra ($R\sim\,$70,000). A data-driven model was trained on the \citet{Brewer2016} spectroscopic parameters using \texttt{The Cannon} \citep{Ness2015}, and this model was then applied to pre-2004 HIRES spectra to provide new measurements. Of the 362 total stars, we only use the 275 with measurements that fall within the ``reliable range'' of 0.0044--18.71 km~s$^{-1}$ as defined by \citet{Rice2020}, and we use their 1$\sigma$ uncertainty estimate of 0.98 km~s$^{-1}$. Values $\leq0.1$~km/s are marked as non-detections.

\item \citet{Luck2017}: Sandiford: \vsini measurements were calculated for 463 stars on the HPIC list using spectra from the Sandiford Cassegrain Echelle Spectrograph ($R\sim\,$60,000) on the McDonald Observatory 2.1\,m telescope. Values were determined by fitting synthetic spectra in a narrow region around 570 nm with MARCS model atmospheres \citep{Gustafsson2008}. The author accounts for $v_\mathrm{mic}$ by constraining the \ion{Fe}{1} abundances to be invariant with equivalent width. Error bars of $1.0$~km/s were adopted and an offset of $-2.2$~km/s was applied based on comparison to \citet{Brewer2016}.

\item \citet{Martinez-Arnaiz2010}: SARG: \vsini measurements were calculated for 128 stars on the HPIC list using spectra from the Spectrografo di Alta Resoluzione Galileo (SARG) spectrograph ($R\sim\,$57,000) at the Telescopio Nazionale Galileo. The rotational velocity was determined by fitting the width of the CCF and removing other broadening contributions by assuming these are governed solely by the $B-V$ color of the star. The authors also account for a spectrograph coupling constant, which is derived for each instrument using slow-rotating stars (i.e., stars assumed to have no detectable rotational broadening). We do not include measurements for which only an upper limit is provided. Error bars of $1.3$~km/s were adopted and an offset of $-3.4$~km/s was applied based on comparison to \citet{Brewer2016}. Non-detections, typically corresponding to \vsini $\lesssim3.3$~km/s, are included as such.

\item \citet{Martinez-Arnaiz2010}: McDonald: \vsini measurements were calculated for 75 stars on the HPIC list using archival spectra ($R\sim\,$50,000) from the 2.7\,m Coud\'e spectrograph at McDonald Observatory taken from \citet{AllendePrieto2004}. The rotational velocity was determined by fitting the width of the CCF and removing other broadening contributions by assuming these are governed solely by the $B-V$ color of the star. The authors also account for a spectrograph coupling constant, which is derived for each instrument using slow-rotating stars (i.e., stars assumed to have no detectable rotational broadening). Error bars of $2.1$~km/s were adopted and an offset of $-1.9$~km/s was applied based on comparison to \citet{Brewer2016}. Non-detections are marked as such.

\item \cite{Ammler-vonEiff2012}: \vsini measurements were calculated for 83 stars on the HPIC list using the least squares deconvolution method as detailed in \citet{Reiners2002,Reiners2003a,Reiners2003b}. The authors used spectra from FEROS ($R\sim\,$48,000) and CES ($R\sim\,$220,000), but the source of the \vsini measurement is not specified in the resulting catalog. We thus rank this catalog based on the lower resolution of FEROS. In the case of duplicate measurements for a single star, we select the measurement with the smaller reported uncertainty. Non-detections, which correspond to \vsini$\lesssim8$~km/s, are marked as such.

\item \cite{Zakhozhay2022}: \vsini measurements were calculated for 64 stars on the HPIC list using spectra from the FEROS spectrograph ($R\sim\,$48,000). Rotational velocities were determined using the ZASPE pipeline \citep{Brahm2017}, which fits the observed spectra using grids of synthetic spectra. $v_\mathrm{mac}$ values were determined using an empirical relation with stellar effective temperature \citep{Valenti2005}, while $v_\mathrm{mic}$ was set to depend on both effective temperature and surface gravity. Uncertainties were determined using an empirical relation, and the authors impose a lower limit on \vsini of 3 km~s$^{-1}$ based on the instrumental spectral resolution, which we include as non-detections.

\item \citet{Martinez-Arnaiz2010}: FEROS: \vsini measurements were calculated for 25 stars on the HPIC list using archival FEROS spectra ($R\sim\,$48,000) from \citet{AllendePrieto2004}. The rotational velocity was determined by fitting the width of the CCF and removing other broadening contributions by assuming these are governed solely by the $B-V$ color of the star. The authors also account for a spectrograph coupling constant, which is derived for each instrument using slow-rotating stars (i.e., stars assumed to have no detectable rotational broadening). There was insufficient overlap with other sources to estimate errors, so we input an error of $2$~km/s. Non-detections are included as such.

\item \citet{Luck2017}: ELODIE:
\vsini measurements were calculated for 259 stars on the HPIC list using archival ELODIE spectra ($R\sim\,$42,000) from \citet{Moultaka2004}. Values were determined by fitting synthetic spectra in a narrow region around 570 nm with MARCS model atmospheres \citep{Gustafsson2008}. The author accounts for $v_\mathrm{mic}$ by constraining the \ion{Fe}{1} abundances to be invariant with equivalent width. Error bars of $1.1$~km/s were adopted and an offset of $-1.1$~km/s was applied based on comparison to \citet{Brewer2016}.

\item \citet{Luck2017}: HET: \vsini measurements were calculated for 53 stars on the HPIC list using spectra from the High Resolution Spectrograph ($R\sim\,$30,000) on the Hobby-Eberly Telsecope. Values were determined by fitting synthetic spectra in a narrow region around 570 nm with MARCS model atmospheres \citep{Gustafsson2008}. The author accounts for $v_\mathrm{mic}$ by constraining the \ion{Fe}{1} abundances to be invariant with equivalent width. Error bars of $0.62$~km/s were adopted based on comparison to \citet{Brewer2016}.

\item \citet{Schroder2009}: \vsini measurements are provided for 338 stars on the HPIC list. The values were derived from $R\sim48,000$ FEROS spectra and $R\sim40,000$ FOCES spectra using a least squares deconvolution; further details provided in the paper and references. Authors note that spectral resolution limits results to $\sim$4~km/s. Error bars of $1.6$~km/s were adopted based on comparison to \citet{Holmberg2007}. Based on comparison to \citet{Brewer2016}, we mark $<6$~km/s as non-detections.

\item \citet{Holmberg2007}: \vsini measurements are calculated for 3248 stars on the HPIC list using medium resolution ($R\sim\,$20,000) spectra taken with the CORAVEL and CfA spectrographs. For the CORAVEL spectra,  \vsini values are calculated using empirical CCF width-\vsini relations calibrated by \citet{Benz1981,Benz1984}. For the CfA spectra, the authors adopt \vsini values form best-fit template spectra. Details of the data and methods are described in \citet{Nordstrom2004}. Error bars of $1.3$~km/s were adopted based on comparison to \citet{Brewer2016}. Non-detections are not discussed and values are only reported to the nearest ones place; we mark values $\leq1$~km/s as non-detections.

\item \cite{Zorec2012}: \vsini measurements are provided for 179 early-type (B and A) stars on the HPIC list. These values are compiled from several different sources with spectral resolutions from $R\sim\,$15,000 to $R\sim\,$28,000. Sources include \citet{Abt1995}, \citet{Abt2002}, and \citet{Levato2004}, all of whom compute \vsini by fitting a Gaussian to observed line profiles, and \citet{Royer2002}, who use the Fourier transform method. Error bars of $9.7$~km/s were adopted based on comparison to \citet{Schroder2009}; these error bars are large due to the early spectral types of the sample and the correspondingly large \vsini. An offset of $-8$~km/s was applied, again derived by comparison to \citet{Schroder2009}.

\item \citet{White2007}: \vsini measurements are provided for 212 stars on the HPIC list. Values were derived from $R\sim\,$16,000 spectra taken with the Palomar 60-in telescope using an empirical CCF width-\vsini relation. The authors note that a lower sensitivity limit of 10~km/s is imposed by the spectrograph resolution. When multiple values are provided for a single star, we adopt the measurement computed from the highest S/N spectrum. Non-detections are included as such.

\end{enumerate}

We note that we excluded values from \citet{Martinez-Arnaiz2010} that originated from FOCES. Comparison between these data and other values in the ARC indicated that these have limited precision.


\subsection{S-index}
\label{sec:s_index}
The \Sindex (also $S$-value or $S_\mathrm{HK}$) was initially defined based on the fluxes in the four channels measured by the Mount Wilson spectrographs as part of a study to monitor stellar \CaII H \& K emission. Two of these channels capture the \CaII H \& K line cores and the other two capture nearby continuum regions. The details of the original \Sindex calculation are described in \citet{Duncan1991}. Measurements of \CaII H \& K emission have since been reported based on observations with dozens of other spectrographs, and for the most part these adopt an \Sindex definition using the same wavelength regions covered by the Mount Wilson channels. Measurements with different instruments have systematic differences, but these differences are usually calibrated and removed by observing stars that overlap with the original Mount Wilson sample. \CaII H \& K emission, and consequently the \Sindex, is time-dependent, so ideally, one would report a time series of measurements over multiple activity cycles. But for the purposes of this catalog, we are interested in a single representative \Sindex measurement for each star. We prioritize catalogs with longer time baselines, as these are more likely to have a median that reflects the typical activity level. We also prioritize more recent data, to capture longer timescale changes, and data taken at higher S/N and higher spectrograph resolution. A few catalogs contained a handful of stars with S indices $>$2---which are unphysical---and thus have not been included in the ARC.

\begin{enumerate}

\item \citet{Baum2022}: \Sindex measurements are taken from four archival sources: (1) Mount Wilson data spanning 1966--2001 from \citet{Radick2018}; (2) pre-upgrade HIRES data from \citet{Wright2004}; (3) post-upgrade HIRES data through 2010 from \citet{Isaacson2010}; and (4) additional archival HIRES data through 2020. All of these archival data were calculated following the standard Mount Wilson calculation from \citet{Duncan1991}. The four data streams are carefully combined with systematic offsets removed, and the median value is reported for each star.

\item \citet{Teklu2025}: \Sindex measurements are calculated using archival HIRES spectra. Measurements are assumed to follow the standard Mount Wilson calculation, as indicated in the catalog that accompanies the manuscript, but the \Sindex calculation is not described in the text of the manuscript. Time series measurements and median values are reported; we adopt the latter. Baselines for each star range from 1 day to 26 years.

\item \citet{Isaacson2024}: \Sindex measurements are calculated using 20 years of HIRES data from the California Legacy Survey \citep{Rosenthal2021}. Measurements follow the standard Mount Wilson calculation and are calibrated to the Mount Wilson scale. Time series measurements and median values are reported; we adopt the latter.

\item \citet{GomesdaSilva2021}: \Sindex measurements are calculated using HARPS data as part of the AMBRE project. Baselines range from 1~d to 15~yr, with a median of 6~yr. Measurements follow the standard Mount Wilson calculation and median values are reported both on an instrument-specific scale and calibrated to the Mount Wilson scale. We adopt the latter.

\item \citet{BoroSaikia2018}: \Sindex measurements are taken from nine archival sources and supplemented with new HARPS observations first presented therein.  For the new HARPS data, \Sindex values are calculated and calibrated to the Mount Wilson scale, but archival measurements are taken at face value. Mean and median values are computed for each stars over all of the compiled measurements; we adopt the latter. This value may not represent the typical stellar activity level in every case, as it is a median of means and medians taken over various baselines, not the median of a time series.

\item \citet{Brewer2016}: \Sindex measurements are calculated on the standard Mount Wilson scale using a single epoch of HIRES data for each star. One star with \RHK $=-1.8$ and $S=540$ was excluded. 

\item \citet{Wright2004}: \Sindex measurements are calculated using archival spectra from HIRES and the Lick-Hamilton echelle spectrograph. Calculations follow the standard Mount Wilson method and are calibrated against the \citet{Duncan1991} sample to place them on the Mount Wilson scale. Median values and time series differential measurements are reported for each star. We adopt the medians. 

\item \citet{Schroder2009}: \Sindex measurements are calculated using archival FEROS ($R\sim48,000$) and FOCES ($R\sim40,000$) spectra. Calculations follow the standard Mount Wilson method and are calibrated against \citet{Baliunas1995} to place them on the Mount Wilson scale. A single value is provided for each star.

\item \citet{White2007}:  \Sindex measurements are calculated using data from the Palomar 60-in spectrograph data, based on a single epoch for each star. Calculations follow the standard Mount Wilson method and are calibrated against \citet{Wright2004} to place them on the Mount Wilson scale. When multiple values are provided for a single star, we adopt the measurement computed from the highest S/N spectrum.

\item \citet{Henry1996}: \Sindex measurements are calculated for southern hemisphere stars using data from the Cassegrain Spectrograph on the 1.5\,m telescope at Cerro Tololo Inter-American Observatory (CTIO). The measurements follow the standard Mount Wilson calculation and are reported both on an instrument-specific scale and calibrated to the Mount Wilson scale. We adopt the latter. The majority of stars had just one epoch. Several had multiple epochs but none had a substantial measurement baseline.

\item \citet{Cincunegui2007}: \Sindex measurements are calculated using data from the 2.15\,m RESOC spectrograph at the CASLEO Argentinean Observatory. Measurements follow the standard Mount Wilson calculation and are calibrated to the Mount Wilson scale. Mean values are reported. Spectra were collected over 7 years, but the baselines vary for different stars.

\item \citet{Zhang2024}: \Sindex measurements are calculated using data from the LAMOST Low-Resolution Spectroscopic Survey. The measurements are calibrated to the Mount Wilson scale using an empirical fit to 65 stars that are shared between the LAMOST and Mount Wilson surveys as described in \citet{Zhang2022}. A single value is reported for each star.

\end{enumerate}


\subsection{Chromospheric Activity Level}
\label{sec:rhk}
Though the \CaII H \& K emission lines do capture chromospheric activity, the \Sindex also includes photospheric contributions. \citet{Noyes1984} defined a new metric, \RHK, usually reported as \logRHK, which isolates the chromospheric component of the \Sindex and is therefore expected to be a more reliable activity proxy. Literature measurements of \RHK typically follow the calculation laid out in \citet{Noyes1984}. Because \RHK is calculated directly from the \Sindex, these two metrics are subject to mostly the same caveats. However, the conversion of \Sindex to \RHK depends on the $B-V$ color of the star, and this relation is only well calibrated on the (approximate) range $0.4 < B-V < 1.0$. Measurements of \RHK for stars outside of this range are not always calibrated to the same scale, even if the S-indices are consistent. In addition, $B-V$ values are not always measured, and there is no single standard approach to infer them. Here, we prioritize catalogs in the same order as for the \Sindex. We note differences in $B-V$ and \RHK calculations.

The lack of standardization in calculating \logRHK\ can lead to systematic uncertainties when comparing activity levels between stars and populations. In Section~\ref{sec:new_activity}, we provide updated calculations for \logRHK\ that provide improved uniformity.

\begin{enumerate}

\item \citet{Teklu2025}: \RHK values are extracted from archival HIRES spectra using a bespoke analysis method rather than via \Sindex transformation. The methods are described in \citet{Perdelwitz2021} and \citet{Perdelwitz2024}. Time series measurements and median values are reported; we adopt the latter. Baselines for each star range from 1 day to 26 years.

\item \citet{Isaacson2024}: \RHK values are calculated and reported based on $B-V$ colors determined using a stellar parameter relation from \citet{Ramirez2005} for stars with \Teff 7000--3870 K. For cooler stars, $B-V$ colors are taken directly from \citet{Rosenthal2021}. Authors use standard the $B-V$ relation from \citet{Noyes1984} for all stars, but urge caution that \RHK for cooler stars may be unreliable.

\item \citet{GomesdaSilva2021}: \RHK values are calculated and reported based on $B-V$ colors taken from Hipparcos and de-reddened. The \Sindex is converted to \RHK using a $B-V$ relation from \citet{Rutten1984}, which extends to $B-V = 1.6$ for main sequence stars.

\item \citet{Meunier2022}: \RHK values are calculated and reported based on $B-V$ colors using $B$ and $V$ magnitudes from SIMBAD. \Sindex measurements are derived from archival HARPS data following standard the Mount Wilson calculation, and conversion of the \Sindex to \RHK follows standard \citet{Noyes1984} relation.

\item \citet{Brewer2016}: The source of the $B-V$ colors used to computed \RHK is not explicitly stated. Conversion of the \Sindex to \RHK follows the standard \citet{Noyes1984} relation, even for stars with $B-V$ colors outside of the calibrated range. One star with \RHK $=-1.8$ and $S=540$ was excluded.

\item \citet{Wright2004}: \RHK values are calculated and reported based on $B-V$ colors taken from Hipparcos. The authors use the standard $B-V$ relation from \citet{Noyes1984} and do not calculate \RHK for stars outside the range $0.44 < B-V < 0.9$.

\item \citet{Schroder2009}: The source of the $B-V$ colors used to computed \RHK is not explicitly stated. Conversion of the \Sindex to \RHK follows the standard \citet{Noyes1984} relation, with extension to redder stars.

\item \citet{White2007}: \RHK values are calculated and reported based on  $B-V$ colors taken from the Tycho catalog and transformed according to \citet{Mamajek2002}. For stars not in Tycho, a relation between spectral type and $B-V$ is used \citep{Johnson1966}. Conversion of the \Sindex to \RHK follows the standard \citet{Noyes1984} relation, even for stars with $B-V$ colors outside of the calibrated range. When multiple values are provided for a single star, we adopt the measurement computed from the highest S/N spectrum.

\item \citet{Zhang2024}: \RHK values are calculated and reported based on $B-V$ colors determined using a stellar parameter relation from \citet{Casagrande2010}. Conversion of the \Sindex to \RHK begins with the standard \citet{Noyes1984} relation but includes additional correction factors.

\item \citet{Hinkel2017}: \RHK values are collected from various literature sources. Averages are taken for stars with multiple reported measurements.

\end{enumerate}


\subsection{Rotation Period}
\label{sec:prot}
In addition to spectroscopic measurements of \vsini, stellar rotation periods (\Prot) can be determined from time-series photometry or from periodic variations in spectroscopic activity indicators, such as \CaII H \& K, \Sindex, and H$\alpha$. As a star rotates, active regions such as spots and faculae will traverse the observed hemisphere, causing variations in the total observed stellar flux over time. While starspots come and go over time, they can persist for several stellar rotation periods \citep{Strassmeier2009}. Rotation periods are measured using a variety of periodogram analysis techniques---including several methods based on the Lomb-Scargle \citep[LS;][]{Lomb1976, Scargle1982,Zechmeister2009} Periodogram for unequally spaced data. Overall, we choose to prioritize catalogs with longer observational baselines since these catalogs are more likely to capture longer rotation periods. However, the time sampling for spectroscopic activity indicators is not as frequent as time-series photometry. Therefore, for \Prot we choose to rank rotation periods obtained from photometric methods higher overall than those measured from spectroscopic activity metrics. 

\begin{enumerate}

\item ASAS-SN \citep{Jayasinghe2018, Christy2022}: The All-Sky Automated Survey for Supernova (ASAS-SN) observed the entire sky every night using 24 telescopes around the world. Stellar light curves were classified using machine learning techniques.  Stars that were classified as rotational variables were selected for inclusion in this work. Rotation periods were measured in a range of 0.05--1000~days using a combination of a Generalized Lomb-Scargle \citep[GLS;][]{Zechmeister2009}, Multi-Harmonic Analysis Of Variance \citep[MHAOV;][]{Schwarzenberg-Czerny1996}, Phase Dispersion Minimization \citep[PDM;][]{Stellingwerf1978}, and Box Least Squares \citep[BLS;][]{Kovacs2002}. A few dozen measurements of periods $>$180 days were excluded since these long periods are not feasible for dwarf stars \citep[with possible exception of the oldest M dwarfs;][]{Newton2019}.

\item KELT \citep{Oelkers2018}: The Kilodegree Extremely Little Telescope (KELT) survey includes 9-years of observations from Winer Observatory in the northern hemisphere and 5-years of observations from South African Astronomical Observatory in the southern hemisphere. Variability periods were measured using a LS periodogram, and known KELT alias periods were removed in order to avoid false positive detections. 

\item \citet{Santos2019}: Rotation periods were measured using a combination of wavelet decomposition \citep{Torrence1998}, auto-correlation function \citep[ACF;][]{McQuillan2013}, and composite spectra \citep{Ceillier2016, Ceillier2017} for K and M stars that were observed during the 4-year \kepler mission \citep{Borucki2010}.

\item VSX \citep{Watson2006}: The Variable Star Index (VSX) is a large collection of variable stars based on submissions from the broader astronomical community, including amateur astronomers. Rotation period measurement methods and total observational baseline varied greatly from source-to-source. Only targets that were classified as rotational variables were included. Rotation periods that were appended by ``:'' were excluded.

\item NGTS \citep{Briegal2022}: The Next Generation Transit Survey (NGTS) is an array of twelve 20-cm telescopes based at ESO’s Paranal Observatory in Chile. The survey was designed to obtain approximately 500 hr of photometry over 250 nights. Periodic variability was measured up to 125 days using a combined generalized ACF and Fast-Fourier Transform \citep{Cooley1969, Harris2020}. 

\item \gaia DR3 \citep{Distefano2023}: The variability catalog from \gaia DR3 included stars that were classified to exhibit rotational modulations. Stellar rotations were measured using a LS periodogram analysis applied to 120-day data segments that contained at least 12 data points. The best rotation period was determined from the mode of the periods measured from all available data segments. 

\item TARS \citep{Boyle2026}: The \tess All-Sky Rotation Survey (TARS) measured rotation periods of stars within 500 pc from the \tess Full-Frame Images. Rotation periods were measured up to 25 days in duration using a LS periodogram and were identified as stellar rotations using random forest classifiers. Rotation periods were measured in each \tess sector separately, then adopted into the catalog based on the outlier-rejected weighted mean of the periods from each sector. 

\item TESS-SVC \citep[][2026 in prep.]{Fetherolf2023}: The \tess Stellar Variability Catalog (TESS-SVC) measured variability periods for stars observed at 2-min cadence by \tess. Variability periods were measured in the range of 0.003--13 days from individual \tess sectors (27-day baseline) using a LS periodogram analysis and ACF. Variability periods that were measured using the ACF are not included since these stars in the TESS-SVC were typically identified as being eclipsing binaries or pulsating stars rather than rotation variables.  

\item \citet{BoroSaikia2018}: Rotation periods were collected from a variety of sources, with data from the Mount Wilson \citep{Messina1999, Saar1999, Petit2008, Hempelmann2016, Olah2016, Gaidos2000} and HARPS \citep{Lovis2011} surveys. 

\item \citet{Olah2016}: Rotation periods were measured from 36 years of Mount Wilson \CaII H \& K data using a short-term Fourier transform \citep[STFT;][]{Kollath2009}. 

\item \citet{SuarezMascareno2015}: Rotation periods were measured using a GLS periodogram analysis of \Sindex and H$\alpha$ activity indicators in HARPS spectra. The observational baselines ranged from 3 to 10 years with tens to hundreds of visits per target. 

\item \citet{Mittag2017}: Rotation periods were measured using a combination of polynomial fitting and a GLS periodogram analysis of \CaII H \& K and \CaII IRT measurements from the TIGRE data \citet{Hempelmann2016}.

\item \citet{Hempelmann2016}: Rotation periods were measured using a LS periodogram analysis of \Sindex measurements from TIGRE spectra. The observational baselines ranged from $\sim$1 month to $\sim$1 year with $\sim$tens of observations per target. Measurements of three stars that were appended by ``?'' are excluded. In cases where two periods are listed (two stars), the first listed is adopted since it is the period with the higher significance peak in the periodogram.

\end{enumerate}

Many of the above catalogs only list stars with measured rotation periods rather than an inclusive list of all stars investigated. However, there are many astrophysical and instrumental reasons for why a star may not have a measured rotation period, including stellar inclination, spot configuration, activity level at the time of observations, instrumental sensitivity, observational baseline, and false positive aliases caused by observational cadences.


\subsection{Activity Cycle Period}
\label{sec:pcyc}
Our Sun exhibits a clear cycle of increased and decreased magnetic activity every 11 years. Measuring activity cycles (\Pcyc) in other stars, especially those similar to our Sun, is much more difficult given the minimum observational baseline required to measure changes in average activity levels. However, these multi-decade long observational baselines are becoming more prevalent and a small sample of activity cycles have been measured through periodic variations in \Sindex values. There are also growing efforts to measure activity cycles using space-based photometry with \kepler and \tess to observe either periodic changes in rotation periods or flux amplitude variations. However, stars with measured activity cycles from these surveys do not overlap with stars in the HPIC \citep{Han2021, Ramsay2024, Chahal2025}. Activity cycle measurements are ranked in our catalog by their observational baselines, with longer baselines having higher priority. 

\begin{enumerate}

\item \citet{Isaacson2024}: Activity cycles were measured by using a LS periodogram search in the range of 100--10,000 days on the \Sindex values obtained by CLS. \citet{Isaacson2010} analyzed a sample of 285 stars with $N_{\rm obs}\geq45$ and detected periodic cycles for 138 of these.

\item \citet{Baum2022}: Activity cycles were measured by using a LS periodogram search on \Sindex values collected from four sources (see \ref{sec:s_index}). A total of 59 stars were analyzed and 14 cycles were detected.

\item \citet{BoroSaikia2018}: Activity cycles were measured using a generalized LS periodogram search on multi-epoch \Sindex measurements from the Mount Wilson and HARPS surveys. From an initial sample of 1131 stars with multiple epochs of observations, 53 cycles were detected.

\item \citet{Baliunas1995}: Activity cycles were measured by using a LS periodogram search on \Sindex values collected from two Mount Wilson datasets \citep{Wilson1978, Vaughan1978}. Periodic cycles were detected for 51 of the 110 stars that were monitored.

\end{enumerate}

We note that while \citet{Isaacson2024} and \citet{BoroSaikia2018} only report detected periodic cycles, \citet{Baum2022} and \citet{Baliunas1995} provide  dispositions for each of the stars analyzed in their studies regardless of whether periodic variations were detected. These dispositions include, e.g., ``flat'' for stars with stable activity levels over a long baseline or ``var'' for detectable but aperiodic variations. We do not include these alternate dispositions in our catalog, however, as the \citet{Baum2022} and \citet{Baliunas1995} samples are comparatively small.


\subsection{Photometric Jitter}
\label{sec:phot_jitter}
While photometric jitter can have a broad range of definitions, it is generally used to encapsulate how the average flux changes over the course of a stellar rotation. This is usually quantified using min-to-max flux variations, average periodic changes in fluxes, or the standard deviation of the light curve. Measurements of photometric jitter are further complicated by stellar activity cycles, since more extreme flux amplitude variations can occur during periods of increased magnetic activity. While the methods for measuring photometric jitter vary greatly between studies, we include those that are available to provide insight into the average activity levels of the potential HWO target stars. In particular, the photometric flux amplitudes are significantly larger than jitter measured using something similar to the standard deviation of the light curve. Due to the inconsistency of how photometric jitter is defined, we simply choose to rank the measurements in the same order as the photometric rotation periods. While the units reported in the literature vary, all units have been converted to ppm in this work. 

\begin{enumerate}

\item ASAS-SN \citep{Jayasinghe2018, Christy2022}: Photometric amplitude variations (mag units, converted to ppm) were measured from the difference in the 2.5th and 97.5th percentile in magnitude. This window in the variations was used in order to avoid outliers. 

\item \citet{Santos2019}: Photometric jitter estimates (ppm units), also considered a photometric proxy for \Sindex ($S$-phot), were estimated from the standard deviation of the \kepler light curve in a baseline window that was 5 times the measured rotation period. 

\item NGTS \citep{Briegal2022}: Photometric amplitude measurements (\%, converted to ppm) were measured from the 5th to 95th percentile flux variations in the light curves. 

\item \gaia DR3 \citep{Distefano2023}: Photometric amplitude variations (mag units, converted to ppm), also referred to as the activity index, were measured from the difference in the 5th and 95th percentile in magnitude.  

\item TARS \citep{Boyle2026}: Photometric amplitude measurements (\%, converted to ppm) were measured based on the median of the amplitudes that were from the sinusoidal models that were best-fit to the light curves from each \tess sector. 

\item TESS-SVC \citep{Fetherolf2023}: Photometric amplitude variations (ppm units) were measured based on a sinusoidal fit to the phase-folded \tess light curves. 

\end{enumerate}


\section{Activity and Rotation Catalog}
\label{sec:catalog}
We present the ARC, which includes activity and rotation measurements compiled from the literature for 7981 stars (at least one property described in Section~\ref{sec:methods}) of the 12,944 stars that are potential targets for the future Habitable Worlds Observatory. All values in the catalog have been compiled from peer-reviewed literature.
For individual properties with multiple literature measurements, we prioritize values based on the ranked order described throughout Section~\ref{sec:methods}. The ARC also includes new \RHK measurements calculated from the adopted \Sindex values with a uniform calibration, which we describe in Section~\ref{sec:new_activity}. For brevity, we only show the column names and descriptions available in the catalog in Table~\ref{tab:arc}, but the complete catalog is available online in a machine-readable format. The most up-to-date version is also provided on GitHub.\footnote{\label{fn:github}\url{https://github.com/ernewton/TSS_ARC}} Completeness of the ARC relative to the prioritized TSS25 target list is reviewed in Section~\ref{sec:completeness}. Verification of target cross-matching between literature measurements and the adopted values in the catalog is presented in Appendix~\ref{sec:appendix}. 

\begin{deluxetable*}{llllccc}
\tablecaption{Column descriptions for the ARC\label{tab:arc}}
\tablehead{\colhead{Column Name}&\colhead{Data Type}&\colhead{Units}&\colhead{Description}&\colhead{$N_1$}&\colhead{$N_2$}&\colhead{$N_3$}}
\startdata
\multicolumn{7}{c}{Columns included from HPIC \citet{Tuchow2025}} \\\hline
star\_name & string & $\cdots$ & Star name& $\ldots$ & $\ldots$ & $\ldots$\\
TSS\_tier & int & $\cdots$ & Target priority tier & 164 & 495 & 12285\\
ra & float & deg & Right Ascension& $\ldots$ & $\ldots$ & $\ldots$\\
dec & float & deg & Declination & $\ldots$ & $\ldots$ & $\ldots$\\
$T_\mathrm{eff}$ & float & K & Effective temperature& 164 & 495 & 12123\\
sy\_dist & float & pc & Gaia distance& 164 & 495 & 12142\\
Bmag & float & mag & Cousins B mag & 164 & 494 & 11772\\
Vmag & float & mag & Cousins V mag & 164 & 495 & 12013\\
Age & float & Gyr & Stellar age & 118 & 335 & 3867\\
\hline\multicolumn{7}{c}{Columns compiled in this work}\\\hline
vsini\_adopt & float & km s$^{-1}$  & Adopted \vsini & 160 & 444 & 3189\\
e\_vsini\_adopt & float &  km s$^{-1}$ & Error on adopted \vsini & 156 & 432 & 3071\\
r\_vsini\_adopt & string & $\cdots$ & Reference for \vsini & $\ldots$ & $\ldots$ & $\ldots$\\
S\_index\_adopt &float &  $-$  & Adopted S index & 147 & 392 & 2028\\
RHK\_adopt & float & $-$ & Reported \RHK & 123 & 308 & 1230\\
prot\_adopt & float & d & Rotation period (photometric and spectroscopic) & 118 & 325 & 5634\\
r\_prot\_adopt & string & $\cdots$ & Source for rotation period & $\ldots$ & $\ldots$ & $\ldots$\\
act\_cycle\_adopt & float & d & Activity cycle period & 31 & 41 & 46\\
r\_act\_cycle\_adopt & float & d & Source for activity cycle period & $\cdots$ & $\cdots$ & $\cdots$\\
phot\_jitter\_adopt & float & ppm  & Photometric jitter & 98 & 302 & 5585\\
r\_phot\_jitter\_adopt & float & ppm  & Source for photometric jitter & $\cdots$ & $\cdots$ & $\cdots$\\
\hline\multicolumn{7}{c}{Columns calculated in this work}\\\hline
RHK\_calculated\_marvin23 & float & $-$ & \RHK recalculated following \citet{Marvin2023} & 147 & 385 & 1996\\
RHK\_calculated\_pyastro & float & $-$ & \RHK recalculated following \texttt{pyastronomy} & 130 & 290 & 1566\\
\multirow{2}{*}{$\tau_c$} & \multirow{2}{*}{float} & \multirow{2}{*}{d}  & Convective overturn timescale calculated from  & \multirow{2}{*}{164} & \multirow{2}{*}{495} & \multirow{2}{*}{12123}\\
& & & $T_\mathrm{eff}$  following \citet{Cranmer2011} & & & \\
\enddata
\tablecomments{The number of measurements for stars in priority Tiers 1, 2, and 3 are $N_1$, $N_2$, and $N_3$, respectively. The contents of this table is available in a machine-readable format with the online journal and through the GitHub.\footref{fn:github}}
\end{deluxetable*}


\subsection{Updated Activity Calculations}
\label{sec:new_activity}
In addition to using the \RHK values reported in the literature, we re-calculate \RHK values from the adopted \Sindex values using two different calibrations. Both methods follow the standard approach, starting from the Mount Wilson \Sindex:
\begin{equation}
R'_{HK} = K \times \sigma_B^{-1} \times 10^{-14} \times C_{cf} \times S - R_\mathrm{phot},
\end{equation}
where $\sigma_B$ is the Stefan-Boltzmann constant, $K$ is a unit conversation factor, $C_\mathrm{cf}$ is a bolometric correction factor, and $R_\mathrm{phot}$ measures the photospheric contribution to the \RHK measurement. 

We first use the default parameters from the \texttt{SMW\_RHK} class in \texttt{PyAstronomy}\footnote{\url{https://github.com/sczesla/PyAstronomy}} \citep{Czesla2019}, which includes the $C_\mathrm{cf}$ from \citet{Rutten1984} and $R_\mathrm{phot}$ from \citet{Noyes1984} that are both based on B-V colors (Johnson-Cousins, same as compiled in \citealt{Tuchow2024}). We limit application to $B-V$ colors from 0.44 to 1.2 and $T_\mathrm{eff}>4000$ K. For the second method, we use the calibration from \citet{Marvin2023}, which  is based on $T_\mathrm{eff}$ rather than B-V color and extends to M dwarfs, making it more widely applicable. 

\begin{figure}[ht!]
    \centering
    \includegraphics[width=\linewidth]{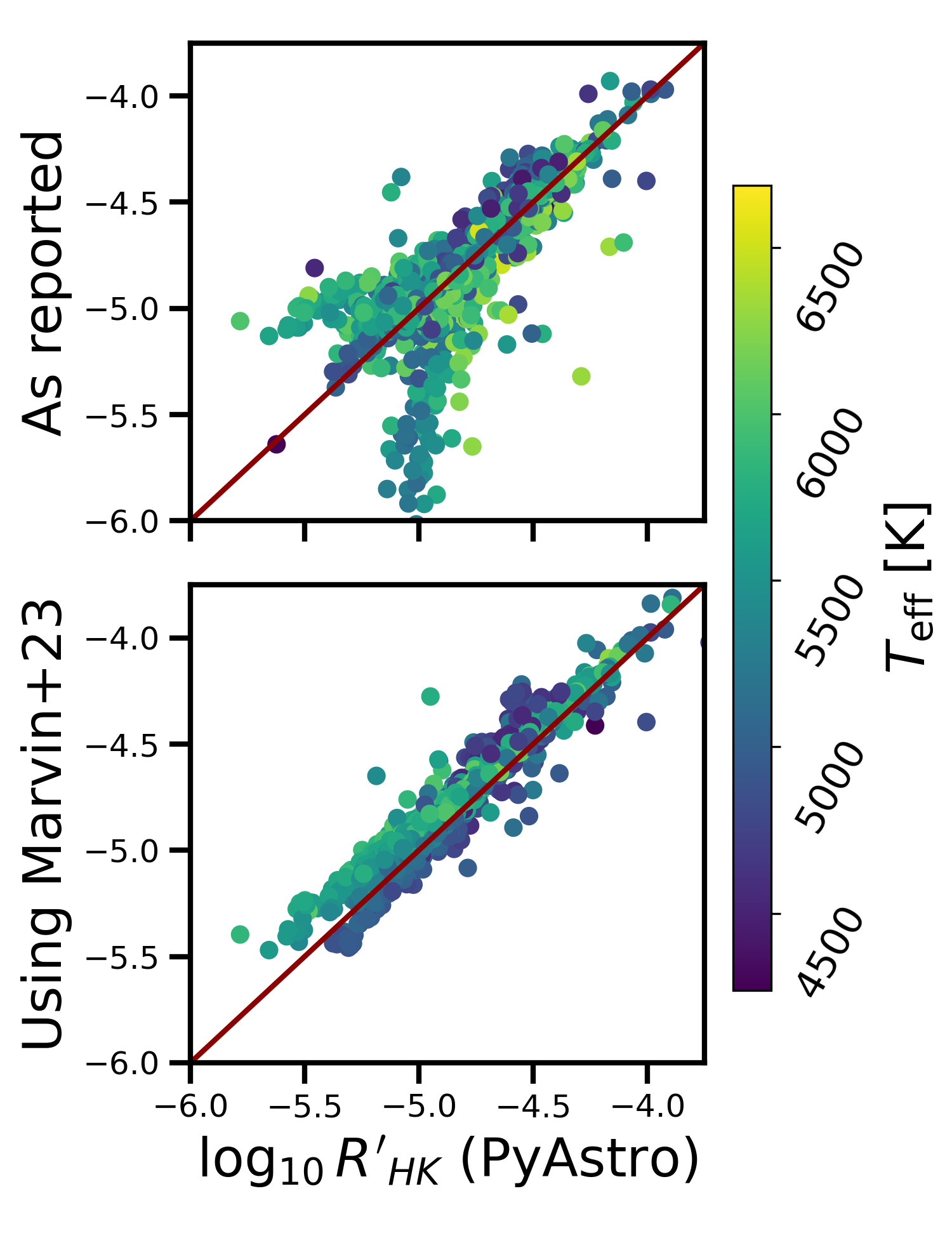}
    \caption{Comparison in \RHK values from the literature (Section~\ref{sec:rhk}) and re-calculated using \citet{Marvin2023}, in comparison to \RHK values recalculated using \texttt{PyAstronomy}.  The points are colored by stellar effective temperatures and the one-to-one relationship is shown by a solid red line. Because of the limits of the calibrations in the method from \texttt{PyAstronomy}, M dwarfs are not included in these plots. }
    \label{fig:verification-cahk}
\end{figure}

The \RHK values calculated by these methods and adopted the literature are shown in Figure~\ref{fig:verification-cahk}. \RHK from \citet{Marvin2023} and \texttt{PyAstronomy} are in good agreement, with slight differences as a function of temperature, while there is a wide scatter in the literature values. The majority of the scatter in the literature values derives from the \RHK calculation (the \Sindex measurements show excellent agreement; see Appendix \ref{sec:appendix}). There are two main contributors: firstly, calculation of the B-V colors and secondly, the conversion from \Sindex\ to \RHK. The calibration from \citet{Marvin2023} circumvents both issues due its use of \Teff\ instead of B-V. While approaches to inferring \Teff\ also vary, surveys such as Gaia provide uniform measurements for many stars. We note that the \Teff\ we use here come from \citet{Tuchow2025} and are compiled from multiple sources.


\subsection{Catalog Completeness}
\label{sec:completeness}
\begin{figure*}[ht!]
    \centering
    \includegraphics[width=\linewidth]{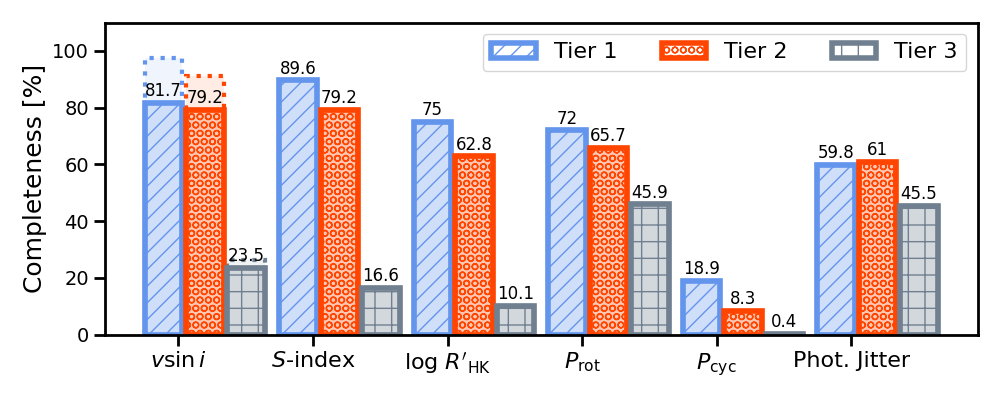}
    \caption{Completeness fractions for each of the stellar parameters included in the ARC. These are broken down by priority tier to highlight the relative sizes of the gaps in our knowledge between the highest priority (Tier 1) and lowest priority (Tier 3) stellar samples. The dashed bars for \vsini indicate the added completeness when including upper limits.}
    \label{fig:completeness}
\end{figure*}

The ARC includes stellar activity and rotation measurements that are currently available for potential HWO target stars that are identified in the HPIC \citep{Tuchow2024}. These stars have since been further divided into three prioritized categories in TSS25 \citep{Tuchow2025}. The highest priority category (Tier 1) consists of the 164 stars on the NASA Exoplanet Exploration Program Mission Star List \citep{Mamajek2024}, which was constructed to identify the stars whose exo-Earth companions, if present, would be the most amenable for atmospheric characterization with future imaging surveys such as the HWO. The second highest priority category (Tier 2) contains 495 stars that could plausibly be observed by the HWO exo-Earth survey while being agnostic to the final telescope architecture \citep{Tuchow2025}. Finally, the lowest priority category (Tier 3) includes the remaining 12,285 stars nearby and bright stars in the HPIC  \citep{Tuchow2024}. The number of stars with measurements for each activity and rotation parameter reported in the ARC is provided in Table~\ref{tab:arc}. The percentage of completeness for each activity and rotation property divided by tier is shown in Figure~\ref{fig:completeness}.

An ideal goal leading up to the launch of the HWO would be to accurately measure and constrain activity and rotation properties of all stars in the Tier 1 and 2 priority categories.
For stars in these tiers, the parameters that can be derived from single-epoch spectroscopy (\vsini, \Sindex, and \RHK) already have high completeness. In particular, \vsini is the most commonly measured activity and rotation metric in Tiers 1 and 2, with measurements compiled in the ARC for more than $\sim$80--90\% of stars depending on whether non-detections (dashed bars). On the other hand, \Prot is only measured for $\sim$60--70\% of the stars in the Tier 1 and 2 priority lists. While \vsini and \Prot both represent measurements of stellar rotation, being able to measure \Prot depends on several astrophysical and instrumental factors, such as stellar inclination, spot configuration, activity level at the time of observations, instrumental sensitivity, observational baseline, and false positive aliases caused by observational cadence. The \Sindex and \logRHK activity metrics are generally well measured across Tiers 1 and 2 (60--90\%), but these measurements represent a snapshot of stellar activity at the time of the observations. Changes in \Sindex and \logRHK are expected to occur throughout a star's activity cycle, but activity cycle measurements are notably absent for the majority of HPIC stars at all tiers ($<$20\%). This completeness fraction would increase if one considers non-cycling stars, but definitive non-detections of cycles are not consistently reported in the literature. Photometric jitter is moderately constrained across the HPIC ($\sim$35--60\%), but this activity metric is inconsistently defined across the literature (see Section~\ref{sec:phot_jitter}), which decreases its utility in characterizing stars in a homogeneous way. 


\section{Discussion}
\label{sec:discussion}
In the following subsections, we investigate relationships between the activity and rotation properties based on the measurements that have been adopted in the ARC. In Section~\ref{sec:rotation}, we specifically compare stellar rotation metrics \vsini and \Prot, and scrutinize their utility for constraining stellar inclinations. In Section~\ref{sec:relationships}, we review relationships between rotation, activity, and stellar ages. We also discuss the relationship between rotation period and activity cycle in Section~\ref{sec:cycle_vs_rot}.


\subsection{Rotation Period vs. Projected Rotational Velocity}
\label{sec:rotation}
\begin{figure*}[ht!]
    \centering
    \includegraphics[width=\linewidth]{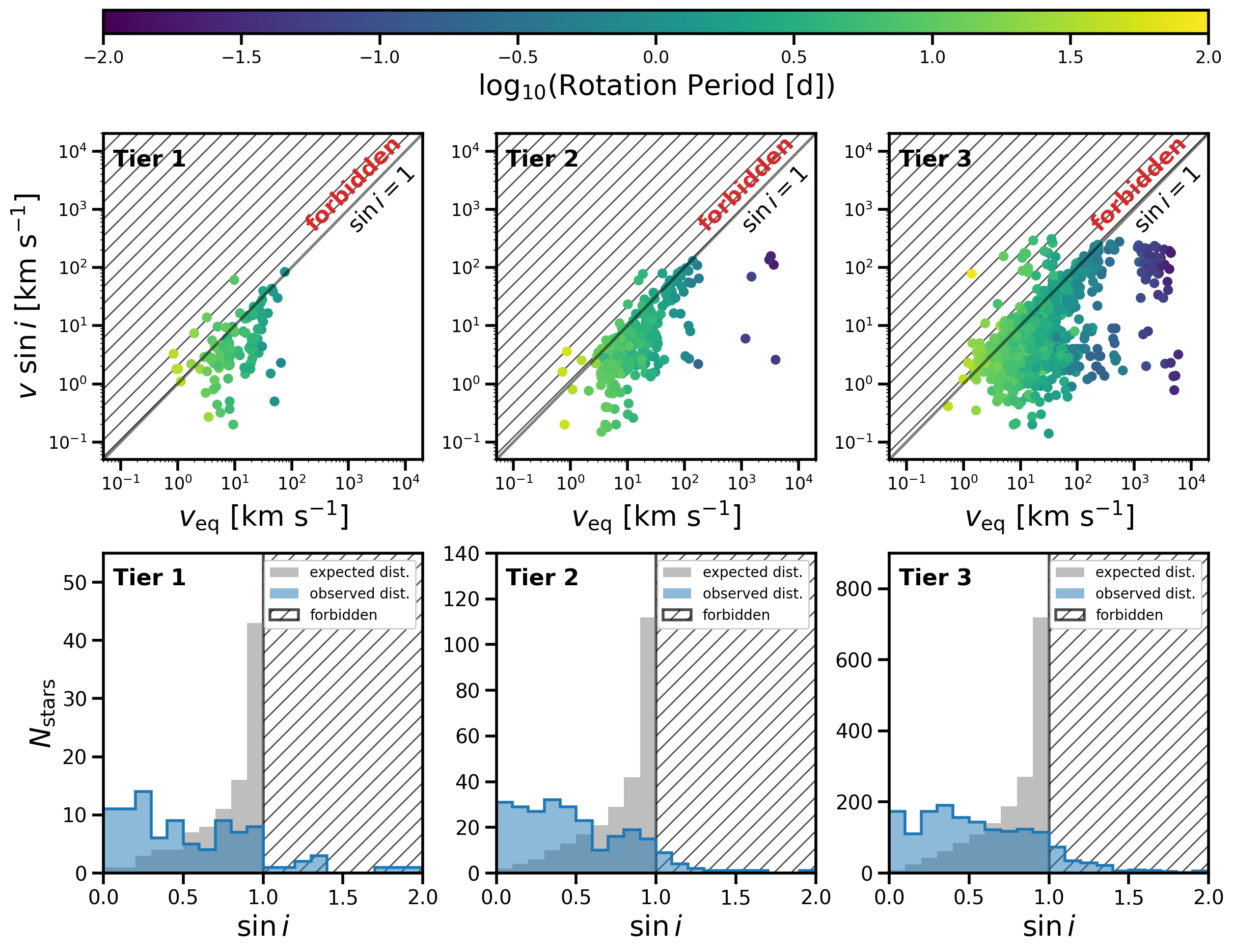}
    \caption{\textbf{Upper panels:} Equatorial rotational velocity ($v_{\rm eq}$) vs. projected rotational velocity (\vsini) for stars in Tiers 1, 2, and 3. The \vsini values are taken directly from the ARC and the $v_{\rm eq}$ values are calculated using \Prot from the ARC and $R_\star$ from the HPIC. Several stars in each tier fall in a region of parameter space that is mathematically forbidden, i.e., requiring $\sin i>1$. \textbf{Lower Panels:} Expected and observed inclination distributions for stars in Tiers 1, 2, and 3. The observed distributions are systematically distinct from the expected distributions. Not only do they predict undefined stellar inclinations, but the results favor face-on orbits ($i=0^\circ$) rather than edge-on orbits ($i=90^\circ$) as one would expect if stellar spin axes were drawn from an isotropic distribution.}
    \label{fig:sini}
\end{figure*}

The catalog constructed herein contains measurements of both \Prot and \vsini for a total of 2177 stars across all priority tiers. This constitutes the largest existing compilation of stars with measurements of both of these rotation metrics. In the case that stellar oblateness and differential rotation are assumed to be negligible, these values should adhere to the following relation:
\begin{equation}\label{eq:incl}
    v \sin i = \frac{2\pi R_\star}{P_{\rm rot}} \sin i,
\end{equation}
such that the inclination of the stellar spin axis, $i$, can be calculated if $R_\star$ is known\footnote{We adopt the standard convention such that $i=90^\circ$ corresponds to an edge-on orbit or a spin axis aligned with the line of sight, and $i=0^\circ$ corresponds to a face-on orbit or a spin axis perpendicular to the line of sight}. Extracting the inclination of the spin axis for such a large sample of stars would be of great interest for a number of exoplanetary and stellar science cases. Notable applications include de-projecting obliquities from Rossiter-McLaughlin measurements to better understand spin-orbit alignments at the population level \citep[e.g.,][]{Rossi2026} and breaking spot coverage degeneracies in the context of transit studies \citep[e.g.,][]{Mori2025}.  To assess whether our catalog will yield reliable measurements of $i$, we use the \Prot and \vsini values from the ARC along with stellar radii as given in the HPIC to calculate the $\sin i$ distribution for stars in each of Tiers 1, 2, and 3. These observed distributions are shown in \autoref{fig:sini} alongside the distribution one would expect if the stellar spin axes were distributed isotropically. For an isotropic distribution, \citet{Masuda2020} show that $\sin i$ should follow
\begin{equation}
    \mathcal{P}_{\sin i}(\sin i) = \frac{\sin i}{\sqrt{1-\sin^2 i}}.
\end{equation}

Not only is it evident that our calculated values are systematically biased, favoring much smaller inclinations than one would expect, but there is also a subset of stars for which we compute $\sin i > 1$ and thus an undefined inclination. We therefore advise that this catalog not be used to compute stellar inclinations without more careful vetting of the source data for individual stars.

To better understand the source of the bias and the forbidden values of $\sin i$, we compute the equatorial velocity $v_{\rm eq}$ as
\begin{equation}
    v_{\rm eq} = \frac{2\pi R_\star}{P_{\rm rot}}
\end{equation}
and compare $v_{\rm eq}$ to \vsini in \autoref{fig:sini}.  Many of the compiled pairs of measurements are consistent with each other, i.e., $v_{\rm eq}\geq v \sin i$. However, some measurement pairs fall in a ``forbidden'' region that requires $\sin i>1$. This will occur if the literature measurements of either \vsini or \Prot exceed the true values. We find the former scenario to be highly likely; as shown in \autoref{fig:compare_vsini}, some sources of \vsini show systematic offsets at \vsini$<10$. This pattern is commonly seen for spectrographs whose lower internal resolutions limit the reliability of broadening measurements for slow rotating stars. It becomes increasingly difficult to accurately measure \vsini and disentangle its effects from those of other line broadening sources (e.g., $v_{\rm mac}$ and $v_{\rm mic}$; see Section~\ref{sec:vsini}) as the total astrophysical broadening decreases or as the size of the spectral resolution element increases \citep{Blanco-Cuaresma2019}. Measurements that were originally reported as upper limits in their source catalogs are not included in \autoref{fig:sini}, but there are likely many instances in which limits are not accurately recorded as such.

The overabundance of low inclinations in the observed distribution (\autoref{fig:sini}) can partly be attributed to known shortcomings of \Prot estimates from \tess data, as described in, e.g., \citet{Fetherolf2023}. The relatively short baseline of individual \tess sectors frequently leads to detections of signals at shorter harmonics of the true rotation period. If some of these stars' rotation periods are in fact longer than the values we adopt for the ARC, the bias in the derived inclination distribution will be somewhat alleviated.


\subsection{Rotation, Activity, and Age Relations}
\label{sec:relationships}

The decline of stellar activity and rotation with age was established in \citet{Wilson1964} and \citet{Skumanich1972}. The relations have been the subject of numerous studies since \citep[e.g.][]{Middelkoop1982, Noyes1984, Rutten1987, Soderblom1993, Randich2000, Wright2004, White2007}, including in some of the sources we compile for the ARC. Although low-mass M dwarfs take a few Gyr to begin their spin down and the associated decline in activity \citep{Newton2016, Pass2024}, most of the stars in the work are solar-type and therefore experience spin down at field ages; in this regime, stellar temperature and age uniquely determine rotation \citep{Barnes2003}. We note exceptions for stalling around $\sim$1 Gyr and weakened braking around $\sim5$ Gyr, depending on mass; see \citet{vanSaders2016, Agueros2018, Curtis2019, Metcalfe2025} for further discussion. The rotation-activity relation is often described as a broken power law, with activity maintaining a saturated level for rapid rotators before declining with increasing rotation. 

\begin{figure}[ht!]
    \centering
    \includegraphics[width=\linewidth]{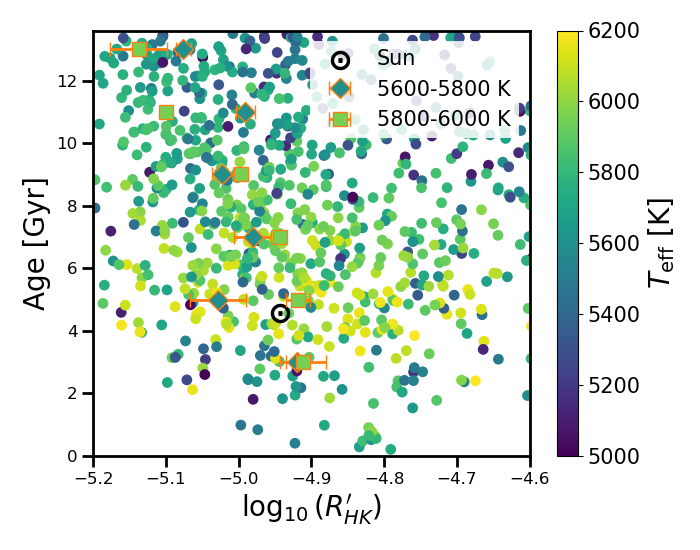}
    \caption{Chromospheric activity \logRHK colored by effective temperature. The gray points represent stars with temperatures less than 5000\,K. The red outlined diamonds and squares show median bins in two bins of effective temperature after the high-activity group is clipped. Ages are derived from \citet{Tuchow2024} and the \logRHK values are updated activity measurements calculated using \texttt{PyAstronomy} (see Section~\ref{sec:new_activity}). Solar values are \logRHK$=-4.9427$ \citep{Egeland2017} and age $=4568.7$ Myr \citep{Piralla2023}. The median activity level declines clearly with age and matches the Sun.}
    \label{fig:act-age}
\end{figure}

\begin{figure}[ht!]
    \centering
    \includegraphics[width=\linewidth]{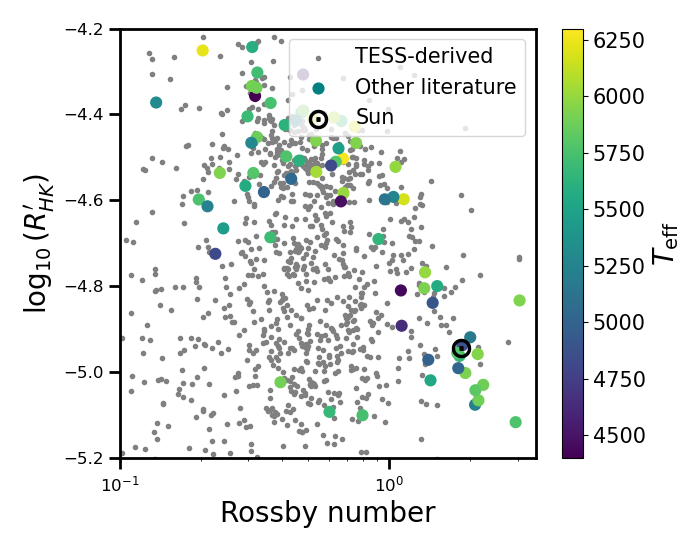}
    \caption{Chromospheric activity \logRHK versus Rossby number, colored by effective temperature. Large Rossby numbers are slower rotators. The gray points represent stars with period measurements from \citet{Fetherolf2023}, which uses \tess\ data and is therefore limited to \Prot~$\lesssim13$~days. The \logRHK values are updated activity measurements calculated using the default \texttt{PyAstronomy} implementation (see Section~\ref{sec:new_activity}). For the colored points, we see the saturated regime at Rossby numbers less than about $0.5$ and the unsaturated regime at larger Rossby numbers/longer periods.}
    \label{fig:rot-act}
\end{figure}

In Figures \ref{fig:act-age} and \ref{fig:rot-act} we show the age-activity and rotation-activity relations based on activity and rotation measurements from the ARC and ages from the HPIC \citep{Tuchow2024}. \citet{Tuchow2024} adopted ages from an imhomogeneous set of reference catalogs (see their Table 4), and they note that the HPIC contains age measurements for only 33.4\% of stars. In the rotation-activity diagram, rotation is presented with a temperature-normalized alternative, the Rossby number. The Rossby number is defined by $P_\mathrm{rot}/\tau$, where $\tau$ is the convection turnover timescale which we calculated using Eqn.~36 from \citet{Cranmer2011}. For improved consistency, we use \RHK re-calculated using \texttt{PyAstronomy} (see Section~\ref{sec:new_activity}). 

The general patterns that we see in Figures~\ref{fig:act-age} and \ref{fig:rot-act} are consistent with previous work. There are two clusters of activity levels \citep[e.g.,][]{Isaacson2024}, separated by the Vaughan-Preston gap. There is one cluster around \RHK~$=-4.5$ (not included in Figure~\ref{fig:act-age}), corresponding to the saturated regime of the rotation-activity relation, and a wider distribution of activity levels at \RHK~$<-4.7$. The gap is typically associated with convergent spin-down at young ages; its presence at a range of older ages in our data could indicate short-period binaries or erroneous ages.

The age-activity relation in Figure~\ref{fig:act-age} can be compared to Fig.~6 in \citet{Mamajek2008} and Fig.~8 in \citet{Soderblom1991}. The former fit a power law relation between $log_{10}(\mathrm{age})$ and \RHK using the mean activity levels of stars in open clusters; as a result, their calibration emphasizes changes over the first Gyr.  The latter fit a Skumanich-style law at field ages. Both fits diverge from our data at field ages, suggesting a need for a revised calibration. K stars show more scatter, likely due to difficulties in age estimations, and are indicated as gray points in Figures \ref{fig:act-age}. 

In the rotation-activity diagram shown in Figure~\ref{fig:rot-act}, we see the two regimes of the rotation-activity diagram. At Rossby numbers less than about 0.5, we see no relation between activity and rotation--the ``saturated'' regime. For larger Rossby numbers (longer periods), activity shows a power law decline with increasing Rossby number as expected for this ``unsaturated'' regime. Significant scatter is seen in the gray dots, which are rotation period measurements from \citet{Fetherolf2023}; this is expected because the TESS-SVC is limited to \Prot$<13$ days, and longer period rotators can be erroneously assigned a short period. 

Recent work has found support for the ``weakened magnetic braking'' scenario first proposed by \citet{vanSaders2016} when analyzing stars from \kepler with rotation periods and asteroseismic ages. In this scenario, stellar spin-down slows significantly around the Rossby number of the Sun, resulting in a discrepancy between rotation-based ages (gyrochronology) and asteroseismic and cluster ages. \citet{Metcalfe2025} showed that this transition is associated with a decrease in X-ray activity and a decrease in inferred stellar wind torque. In Figure~\ref{fig:rot-act}, the Sun's position near the extreme of low activity and large Rossby number is consistent with weakened magnetic breaking. That the age-activity relation apparently persists may relate to the changes in the dynamo structure, as discussed in \citet{Metcalfe2025}.


\subsection{Activity Cycle vs. Rotation}
\label{sec:cycle_vs_rot}
\begin{figure}[ht!]
    \centering
    \includegraphics[width=\linewidth]{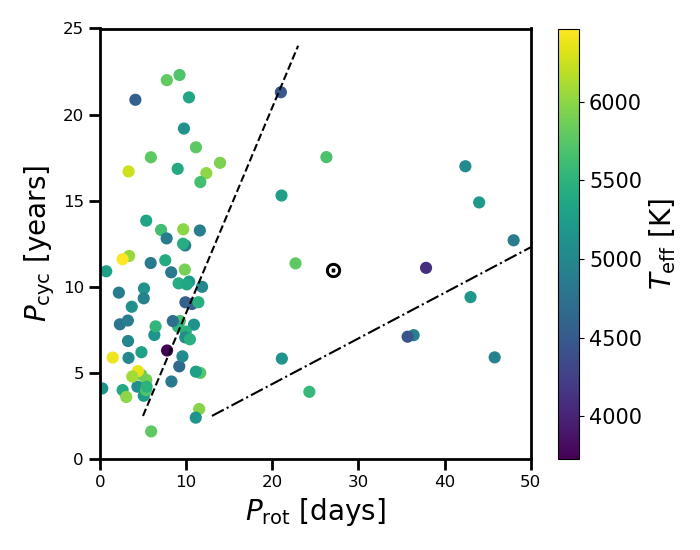}
    \caption{Stellar activity cycle period versus rotation period with points colored by effective temperature for stars with these measurements available in the ARC. The active and inactive branch relationships from \citet{Bohm-Vitense2007} are shown by the dashed and dash-dotted lines, respectively. For reference, the Sun is also indicated by $\odot$. The scatter relative to the active and inactive relationships is consistent with previous findings.}
    \label{fig:cycle_rot}
\end{figure}

\begin{figure}[ht!]
    \centering
    \includegraphics[width=\linewidth]{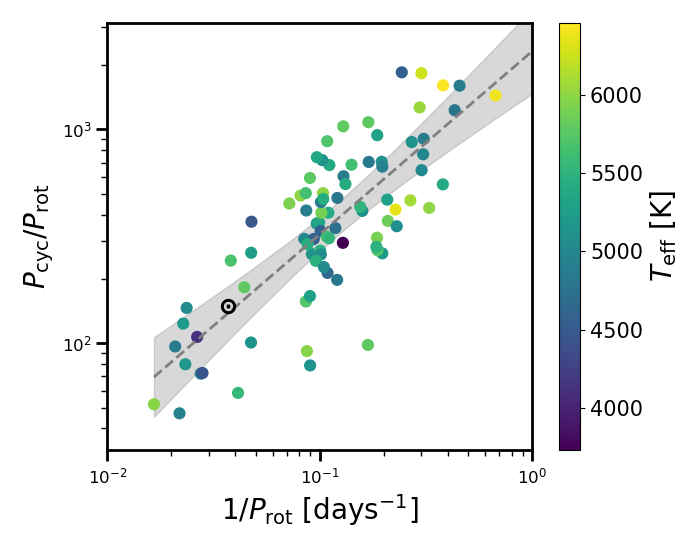}
    \caption{Ratio of stellar activity cycles to rotation period versus rotation period with points colored by effective temperature. The best-fit linear relationship and its 3$\sigma$ uncertainty are shown by the dashed gray line and gray shaded region, respectively. For reference, the Sun is also indicated by $\odot$. The best-fit linear relationship has a slope of $0.856\pm0.016$, which is slightly shallower than a unity slope. This suggests that there may be a slight correlation between activity cycle period and rotation period.}
    \label{fig:cycle_ratio}
\end{figure}

\begin{figure}[ht!]
    \centering
    \includegraphics[width=\linewidth]{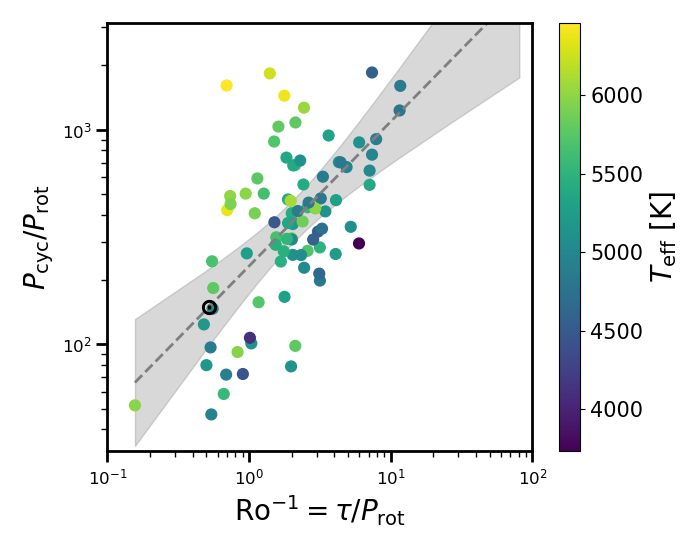}
    \caption{Ratio of stellar activity cycles to rotation period versus the inverse of the Rossby number with points colored by effective temperature. The best-fit linear relationship and its 3$\sigma$ uncertainty are shown by the dashed gray line and gray shaded region, respectively. For reference, the Sun is also indicated by $\odot$. The best-fit linear relationship has a slope of $0.674\pm0.014$, which is shallower than a unity slope. While not as strong as the case for a unity slope, there may be some correlation between the activity cycle period and the convection turnover timescale.}
    \label{fig:cycle_rossby}
\end{figure}

Activity cycles are the most incomplete measured property collected from the literature in the ARC (see Section~\ref{sec:completeness}), but they are important to understand in preparation for future direct imaging missions, such as the HWO. We investigated the relationship between activity cycle and rotation period in Figures~\ref{fig:cycle_rot}--\ref{fig:cycle_rossby}. Similar to previous studies \citep[e.g.,][]{Bohm-Vitense2007, Metcalfe2016, BoroSaikia2018, Han2021, Isaacson2024, Chahal2025}, in Figure~\ref{fig:cycle_rot} we find that stars generally fall near either the active (dashed line) or inactive (dash-dotted line) sequences that were identified by \citet{Bohm-Vitense2007}. We also found that the best-fit linear relationship between \Pcyc/\Prot and 1/\Prot has a slope of $0.856\pm0.016$ that is slightly shallower than a unity slope, where a unity slope would suggest that there is not a direct correlation between activity cycle period and rotation period \citep{SuarezMascareno2016, Chahal2025}. Furthermore, the best-fit linear relationship between \Pcyc/\Prot and the Rossby number has a slope of $0.674\pm0.014$ that is shallower than unity, where a unity slope would imply that stellar activity cycle periods are proportional to convection turnover timescales \citep{Irving2023, Chahal2025}. Slopes shallower than unity for both Figures~\ref{fig:cycle_ratio} and \ref{fig:cycle_rossby} suggest that stellar activity cycle periods could be slightly correlated with the stars' rotation periods and convection turnover timescales. Shallower slopes for these relationships have also been measured in previous studies and have been thought to be due to the spectral type or stage of evolution of the sample, or possibly the complexity of the observed activity cycle \citep[see also][]{Baliunas1996, Olah2009, Olah2016, SuarezMascareno2016, Chahal2025}. Expanding the number of stars with measured activity cycles will enable further studies into the relationship between stellar activity cycles and their rotations, and their connection to stellar dynamo theory. 


\section{Ongoing and Future Efforts}
\label{sec:future}
In addition to continued activity monitoring of HWO target stars with both spectroscopic and photometric facilities, as detailed below, progress needs to be made in our understanding of and ability to mitigate activity contributions to exoplanet characterization efforts. Within the EPRV community, substantial effort has gone into characterizing activity-induced stellar RV variations \citep[see][and references within]{Burt2025}, including recent work on these effects at the spectral level \citep[e.g.,][]{Artigau2022, Gilbertson2024, Siegel2024}. In order to increase the detection of transiting exoplanets, efforts have also been made towards precisely modeling photometric flux contributions from stellar activity \citep[e.g.,][]{Ambikasaran2015, Beard2022, Holcomb2022}. Mitigation techniques for stellar contamination in transmission spectroscopy measurements are also being developed \citep[e.g.,][]{Pinhas2018, Wakeford2019, Rackham2024, Barclay2025}, though much of the focus has been on JWST observations in the near- and mid-infrared, and thus on M-dwarf stellar activity. Research in this vein, with an emphasis on HWO target stars, must continue to be supported if we aim to understand how activity will affect reflected light spectroscopy with HWO.


\subsection{Spectroscopic Activity Monitoring}
Ongoing \citep[e.g.,][]{Motalebi2015, Brewer2020, Gupta2021, Paredes2021} and future \citep[e.g.,][]{Hall2018, Sturmer2024} RV searches for exoplanets and stellar and substellar companions will continue to provide rich activity time series data sets for many Tier 1 HWO target stars. However, due to the time-intensive, single-target nature of precise RV measurements and the need for high cadence observations to robustly sample stellar variability across numerous time scales, not all potential HWO target stars can be monitored by a given survey. Resources need to be allocated to activity monitoring for stars that are not otherwise receiving attention; facilities that are not capable of delivering high precision RV measurements will be particularly suitable for such monitoring programs, as they will not be as sought after for exoplanet searches. Activity monitoring programs can also make use of spectrographs with narrow bandpasses, as in the original Mount Wilson survey, because measurements can be made using a limited number of lines. Other ongoing activity monitoring programs include the HKalpha project at CASLEO observatory \citep{Cincunegui2007,Buccino2024}, which has been collecting data for 25 years.

Future data will inform improved measurements of stellar rotation periods and activity cycles. Newly acquired data from instruments with spectral resolution of close to $R\sim100,000$ (e.g., CARMENES, CHIRON, NEID, EXPRES, KPF)  and approaching and exceeding $R\sim200,000$ (ESPRESSO, PEPSI, iLocater) can also be used to measure previously inaccessible \vsini values at or below the km~s$^{-1}$ level. A targeted survey of Tier 1 HWO stars with one or more of these instruments would be highly valuable and also reasonably tractable over several observing semesters. 


\subsection{Photometric Activity Monitoring}
\tess recently completed its 7th year of observations, and thus far has monitored nearly the entire sky, with three extended visits surveying the northern and southern continuous viewing zones. Year 8 observations are currently underway through September 2026. The \tess mission plans to continue observations until the proposed extended missions are no longer supported or operations are no longer possible. As the baseline of \tess observations continues to grow, stellar activity cycles may become increasingly feasible to constrain using \tess photometry. Not only will the fraction of variable stars increase with photometric precision \citep{Ciardi2011, Basri2011, Briegal2022, Fetherolf2023}, but also observational baseline as quiet stars progress through their activity cycles into periods of increased activity. 

In the landscape of upcoming photometric surveys, the ground-based Rubin Observatory \citep{Ivezic2019} will survey 18,000 deg$^2$ of the sky to a depth of $r \sim 27.5$, which will include millions of stars. While Rubin will only take two short 15-s exposures of a given region of the sky every few nights, these observations may be able to constrain average changes in magnetic activity on long timescales, including potentially measuring long-period rotations. However, many potential HWO target stars are nearby and bright, such that they are likely to be saturated in the Rubin observations. The upcoming Roman space telescope \citep{Kasdin2020, Mosby2020} includes an extended time-series survey of the Galactic bulge, which will be incredibly difficult territory for constraining properties of individual stars due to the extreme crowding. However, there will be inevitable advancements in photometric data analysis methods which will further support efforts to precisely model stellar activity noise. The upcoming Plato space mission \citep{Rauer2025}, meanwhile, will take a similar approach to \kepler by continuously staring at a region of the sky that overlaps with the \tess southern continuous viewing zone. This strategy will likely only cover a small number of potential HWO stars, but will be able to further extend the photometric light curve baselines of these stars and will continue to observe these stars while \tess is observing other sections of the sky.  


\section{Summary}
\label{sec:summary}
Understanding and constraining stellar magnetic activity is important for interpreting observed planetary atmospheres with future direct imaging missions, such as the HWO. Stellar activity can mimic or hide planetary signatures, and can affect our ability to interpret spectra that includes contributions from both the star and the planet. In this work, we aimed to assess our current understanding of stellar activity and rotation in preparation for HWO and other future direct imaging missions. 

In Section~\ref{sec:current_best}, we reviewed current state-of-the-art efforts for measuring activity and rotation properties of stars and mitigating the effects of magnetic activity on measuring planetary properties. We then collected and ranked activity and rotation measurements of potential HWO target stars from the peer-reviewed literature and in Section~\ref{sec:methods} and presented the ARC in Section~\ref{sec:catalog}. The ARC includes measurements for stellar rotations (\vsini and \Prot), magnetic activity (\Sindex and \RHK), and activity cycles (\Pcyc) for stars that may potentially be observed with HWO \citep{Tuchow2025}. Based on the \Sindex values adopted from the literature, we also uniformly calculated updated \RHK values (see Section~\ref{sec:new_activity}). 

We assessed the completeness of measurements of stellar activity and rotation measurements for potential HWO stars in Section~\ref{sec:completeness}. A summary of the completeness is shown in Figure~\ref{fig:completeness}, but in brief we found that projected rotational velocity of stars is the best known metric ($>$90\% Tier 1 and 2 priority stars have measurements) and stellar activity cycles is the most poorly constrained for all potential HWO stars ($<$20\% of stars have measurements). In Section~\ref{sec:rotation}, we investigated the potential for using the \vsini and \Prot measurements from the ARC to infer stellar inclinations. However, inclinations derived in this way were found to be unreliable, such that they require more careful vetting system-by-system. We also reviewed the relationships between rotation, activity, and age in Section~\ref{sec:relationships}, and the relationship between activity cycles and rotation periods in Section~\ref{sec:cycle_vs_rot}.

In Section~\ref{sec:future}, we discussed known ongoing and future efforts for monitoring stellar activity. In particular, there is a critical need to understand how magnetic activity changes over long timescales. Given the long baselines required for measuring magnetic activity cycles, observations that monitor stellar activity need to start now such that these measurements are known prior to planned future missions, including the HWO. Exoplanet properties and their atmospheres are only as well known as their host stars---including our understanding of the star's magnetic activity. Overall, constraining magnetic activity of potential target stars for HWO and other future direct imaging missions is critical for ensuring their success. 


\begin{acknowledgments}
T.F. acknowledges support from an appointment through the NASA Postdoctoral Program at the NASA Astrobiology Center, administered by Oak Ridge Associated Universities under contract with NASA. 
J.A.C. acknowledges financial support from the Spanish Agencia Estatal de Investigaci\'on (AEI/10.13039/501100011033) of the Ministerio de Ciencia, Innovaci\'on y Universidades and the European Regional Development Fund ``A way of making Europe'' through project PID2022-137241NB-C42.
S.C.G. acknowledges financial support from NASA through grant XRP22-2-0187 and the 1.5m SMARTS Telescope Fellowship.
Part of this research was carried out at the Jet Propulsion Laboratory, California Institute of Technology, under a contract with the National Aeronautics and Space Administration (NASA).

This publication is a direct product of the HWO Target Stars and Systems sub-Working Group. The results and conclusions benefited greatly from group-wide discussions and analysis, as well as specific input from the Catalogs \& Databases Task Group, Fundamental Properties Task Group, Activity \& Rotation Task Group, High Energy Emission Task Group, Multiple Stars Task Group, and EPRV Task Group.

\end{acknowledgments}

\software{Astropy \citep{AstropyCollaboration2013, AstropyCollaboration2018},
          Astroquery \citep{Ginsburg2019},
          Matplotlib \citep{Hunter2007},
          NumPy \citep{Harris2020}, 
          SciPy \citep{Virtanen2020}
          }


\appendix 
\restartappendixnumbering


\section{Appendix information}
\label{sec:appendix}
\begin{figure}[ht!]
    \centering
    \includegraphics[width=0.8\linewidth]{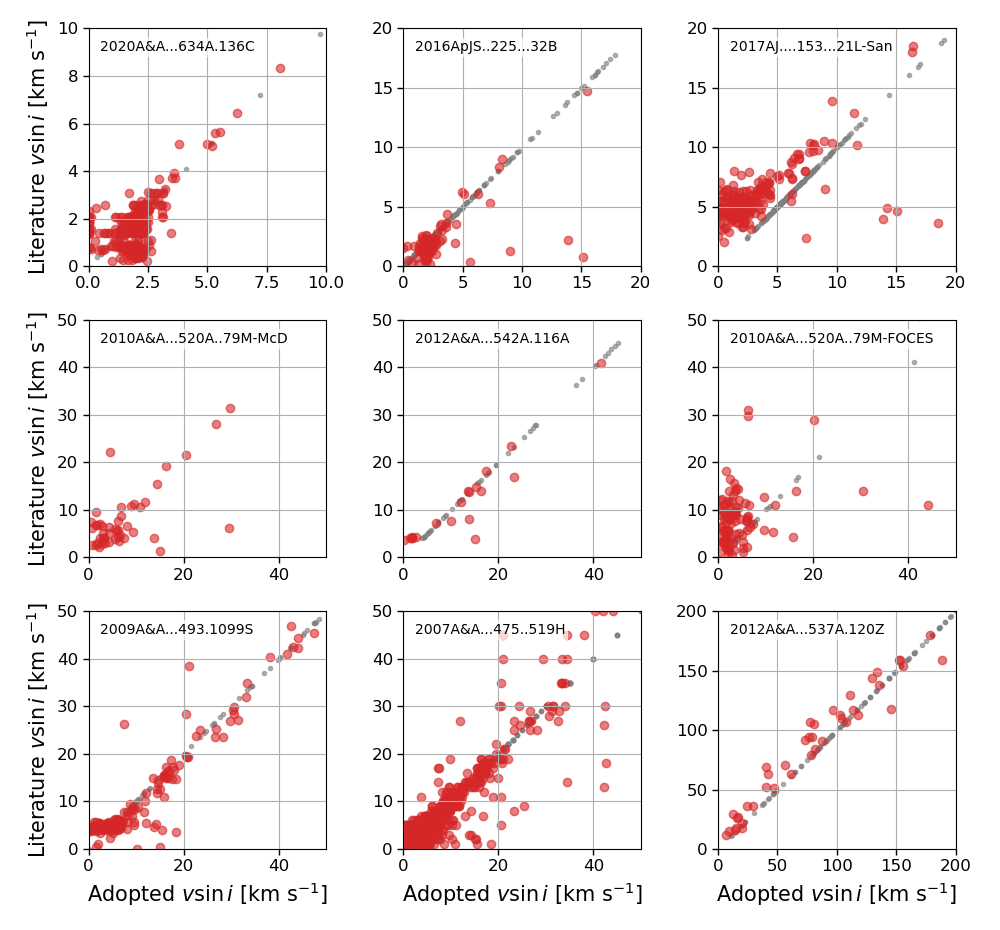}
    \caption{Literature \vsini values versus the \vsini value adopted in the ARC. The bibcode reference is provided in the top left corner of each panel. Gray points represent values that were adopted from the given catalog into the ARC.}
    \label{fig:compare_vsini}
\end{figure}

\begin{figure}[ht!]
    \centering
    \includegraphics[width=0.8\linewidth]{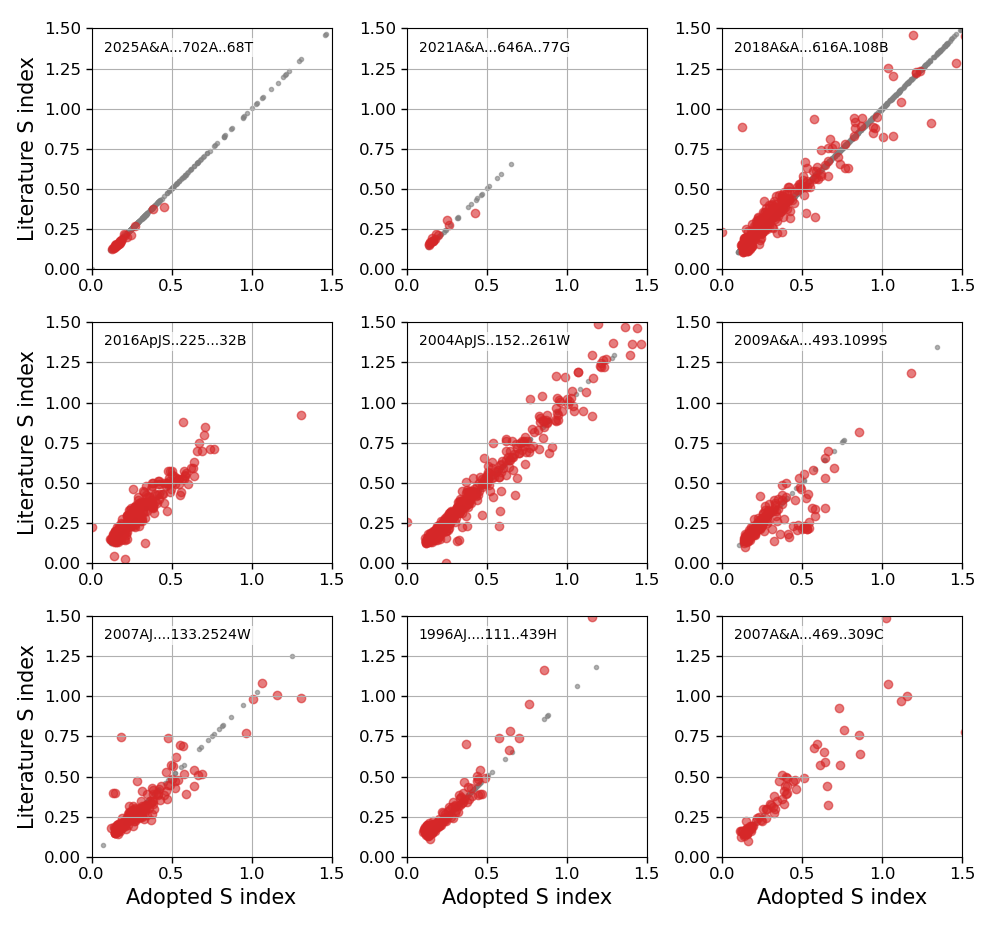}
    \caption{Same as Figure~\ref{fig:compare_sval}, but for \Sindex values.}
    \label{fig:compare_sval}
\end{figure}

\begin{figure}[ht!]
    \centering
    \includegraphics[width=0.5\linewidth]{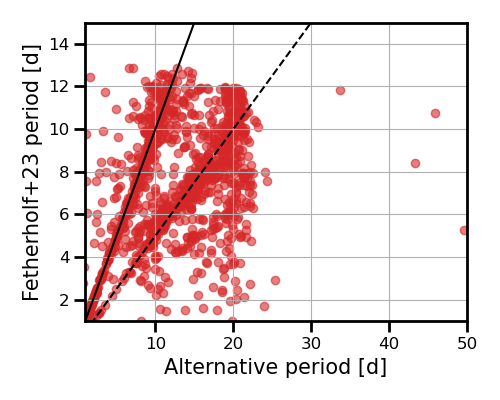}
    \caption{\Prot values from \citet{Fetherolf2023} versus the \Prot value adopted in the ARC. The one-to-one and two-to-one relationships are indicated by the black solid and dashed lines, respectively.}
    \label{fig:compare_period}
\end{figure}

To investigate the validity of our target cross-matching, we compared the values extracted from the literature catalogs against the values adopted in the ARC (see Figures~\ref{fig:compare_vsini}--\ref{fig:compare_period}). For brevity, we only show comparisons from the highest ranked catalogs for \vsini and \Sindex, with the gray points representing measurements adopted in the ARC. In the top left corner of each panel, we provide the bibcode to the literature catalog being compared to the adopted values, with some including an additional identifying tag appended if multiple facilities were presented in the same catalog. Details regarding the specific measurements from each literature catalog can be found in the relevant subsection of Section~\ref{sec:methods}. The adopted \Prot values are only compared to \citet{Fetherolf2023} since this catalog contains a majority of the non-adopted rotation periods. Many of the adopted periods are consistent with being twice what was measured in \citet{Fetherolf2023}, which is likely due to the 13-day limit on their rotation measurements. Literature comparison figures are not provided for \Pcyc due to the lack of measurements of activity cycles overall available, and photometric jitter due to the lack of consistency between measurement methods (see Section~\ref{sec:phot_jitter}). The adopted \RHK values are alternatively compared in Figure~\ref{fig:verification-cahk} to the updated \RHK values calculated and presented in Section~\ref{sec:new_activity}.


\end{document}